\documentclass[useAMS,usenatbib,usegraphicx]{mn2e}

\newcommand{\kms}{\textrm{km~s$^{-1}$}}

\newcommand{\lsolar}{L$_{\odot}$}
\newcommand{\msolar}{M$_{\odot}$}

\newcommand{\ion}[2]{#1$\;${\small{#2}}\relax}

\usepackage{subfigure}
\usepackage{multirow}
\usepackage{amssymb}
\usepackage[dvips]{color}
\definecolor{Mygrey}{gray}{0.6}

\title[Type Ia-CSM SNe 2012ca and 2013dn]{On the Nature of Type Ia-CSM Supernovae: \\Optical and Near-Infrared Spectra of SN 2012ca and SN 2013dn}
\author[O. D. Fox et al.]{Ori D. Fox$^{1,2}$, Jeffrey M. Silverman$^{3}$, Alexei V. Filippenko$^{1}$, Jon Mauerhan$^{1}$, \newauthor Juliette Becker$^{4}$, H. Jacob Borish$^{5}$, S. Bradley Cenko$^{6,7}$, Kelsey I. Clubb$^{1}$, Melissa Graham$^{1}$, \newauthor Eric Hsiao$^{8,9}$, Patrick L. Kelly$^{1}$, William H. Lee$^{10}$, G. H. Marion$^{3}$, Dan Milisavljevic$^{11}$, \newauthor Jerod Parrent$^{11}$, Isaac Shivvers$^{1}$, Michael Skrutskie$^{5}$, Nathan Smith$^{12}$, John Wilson$^{5}$, \newauthor and Weikang Zheng$^{1}$\\
$^{1}$Department of Astronomy, University of California, Berkeley, CA 94720-3411, USA.\\
$^{2}$ofox@berkeley.edu.\\
$^{3}$Department of Astronomy, University of Texas at Austin, Austin, TX 78712, USA.\\
$^{4}$Cahill Center for Astrophysics, California Institute of Technology, Pasadena, CA 91125, USA.\\
$^{5}$Department of Astronomy, University of Virginia, P.O. Box 400325, Charlottesville, VA 22904-4325, USA.\\
$^{6}$Astrophysics Science Division, NASA Goddard Space Flight Center, MC 661, Greenbelt, MD 20771, USA.\\
$^{7}$Joint Space-Science Institute, University of Maryland, College Park, MD 20742, USA.\\
$^{8}$Carnegie Institution of Washington, Las Campanas Observatory, Colina El Pino, Casilla 601, Chile.\\
$^{9}$Department of Physics and Astronomy, Aarhus University, Ny Munkegade, DK-8000 Aarhus C, Denmark.\\
$^{10}$Instituto de Astronom\'ia, Universidad Nacional Aut\'onoma de M\'exico, Apartado Postal 70-264, 04510 M\'exico D.F., Mexico.\\
$^{11}$Harvard-Smithsonian Center for Astrophysics, 60 Garden St., Cambridge, MA 02138, USA.\\
$^{12}$Steward Observatory, University of Arizona, 933 North Cherry Avenue, Tucson, AZ 85721, USA.}
\begin{document}

\maketitle
\begin{abstract}

A growing subset of Type Ia supernovae (SNe~Ia) show evidence for unexpected interaction with a dense circumstellar medium (SNe~Ia-CSM).  The precise nature of the progenitor, however, remains debated owing to spectral ambiguities arising from a strong contribution from the CSM interaction.  Late-time spectra offer potential insight if the post-shock cold, dense shell becomes sufficiently thin and/or the ejecta begin to cross the reverse shock.  To date, few high-quality spectra of this kind exist.  Here we report on the late-time optical and infrared spectra of the SNe~Ia-CSM 2012ca and 2013dn.  These SNe~Ia-CSM spectra exhibit low [\ion{Fe}{III}]/[\ion{Fe}{II}] ratios and strong [\ion{Ca}{II}] at late epochs.  Such characteristics are reminiscent of the super-Chandrasekhar-mass (SC) candidate SN 2009dc, for which these features suggested a low-ionisation state due to high densities, although the broad Fe features admittedly show similarities to the blue ``quasi-continuum'' observed in some core-collapse SNe Ibn and IIn.  Neither SN 2012ca nor any of the other SNe~Ia-CSM show evidence for broad oxygen, carbon, or magnesium in their spectra.  Similar to the interacting Type IIn SN 2005ip, a number of high-ionisation lines are identified in SN 2012ca, including [\ion{S}{III}], [\ion{Ar}{III}], [\ion{Ar}{X}], [\ion{Fe}{VIII}], [\ion{Fe}{X}], and possibly [\ion{Fe}{X}I].  The total bolometric energy output does not exceed $10^{51}$~erg, but does require a large kinetic-to-radiative conversion efficiency.  All of these observations taken together suggest that SNe~Ia-CSM are more consistent with a thermonuclear explosion than a core-collapse event, although detailed radiative transfer models are certainly necessary to confirm these results.

\end{abstract}


\section{Introduction}

Type Ia supernovae (SNe~Ia; see \citealt{filippenko97} for a review) are attributed to the thermonuclear explosion of a C/O white dwarf (WD) primary star that approaches the Chandrasekhar limit by accreting material from a companion star.  While the nature of the companion remains somewhat ambiguous, recent arguments suggest that a WD companion (i.e., a double-degenerate progenitor) is more commonplace than the single-degenerate scenario (see \citealt{maoz13} for a review).  Given that these highly evolved stars have long since shed their outer envelopes, the surrounding circumstellar medium (CSM) is expected to have a relatively low density.  In fact, the lack of a significant CSM is one of the primary reasons SNe~Ia can be used as precise cosmological distance indicators \citep[e.g.,][]{phillips93}.

A growing number of SNe~Ia, however, show evidence of interaction with a dense CSM during the first year post-explosion, and sometimes longer \citep[][and references within]{silverman13b}.  Classified as SNe~Ia-CSM, the spectra have Type Ia features (e.g., \ion{S}{II} and \ion{Si}{II} absorption lines) near maximum light that are weaker than usual, most likely diluted by a strong continuum \citep[e.g.,][]{leloudas13}.  The spectra also have relatively narrow hydrogen emission lines, which arise from the dense and slowly moving CSM formed by pre-SN mass loss that is more typically associated with core-collapse SNe~IIn \citep{schlegel90,filippenko97}.  These dense environments suggest that either (a) single-degenerate progenitor scenarios are responsible for these explosions, or (b) the exploding star was not a thermonuclear explosion of a white dwarf at all.


The SN~Ia-CSM progenitor explosion mechanism (i.e., thermonuclear versus core collapse) remains debated in the literature \citep[e.g.,][]{inserra14}.  For example, the two most well-studied SN~Ia-CSM templates (SNe 2002ic and PTF11kx) were classified as SNe~Ia at early times, showing a resemblance to the overluminous SN 1991T  \citep[][]{hamuy03b, deng04, wood-vasey04, dilday12,silverman13a}.  Compared to SNe IIn, weaker and narrower He, H$\beta$, and O lines in SNe~Ia-CSM further support arguments for a thermonuclear origin \citep{silverman13b}.  For this case of a thermonuclear explosion, \citet{hamuy03b} propose the dense CSM may be attributed to an evolved secondary star (i.e., single-degenerate binary progenitor).  

Alternatively, other SNe~Ia-CSM show less obvious S or Si at early times (e.g., SNe 1997cy and 2005gj), but are only classified as SNe~Ia-CSM because later spectra exhibit iron features and/or match SNe 2002ic and PTF11kx very well \citep[e.g.,][]{germany00,aldering06}.  \citet{benetti06} argue for a core-collapse origin instead, given that some SNe Ic (e.g., SN 2004aw) can be confused for SNe~Ia at early phases if considering only similarities of 6300~\AA\ features most often attributed to \ion{Si}{II} $\lambda$6355.  Furthermore, \citet{inserra14} make line identifications of intermediate/heavy elements in the spectra of  the SN~Ia-CSM 2012ca that are consistent with a core-collapse explosion of an evolved massive star.  The ambiguity is compounded by the fact that the available catalogs of SN~Ia spectra already suggest a number of degeneracies in the spectroscopic classification process (see \citealt{parrent14} for a review).  Furthermore, spectroscopic models of thermonuclear SNe with CSM interaction (CSI) have not yet been constructed. Owing to the strong CSI and underlying continuum, even post-photospheric phase spectra may offer little evidence about the ejecta composition to connect with the pre-maximum spectral type and the related progenitor system \citep{leloudas13}. 

SN 2012ca stands out as being one of the most nearby SNe~Ia-CSM (79 Mpc; see Table \ref{tab_100mpc}), allowing for high-resolution data having a high signal-to-noise ratio (S/N), even at late times after the CSI has faded and the ejecta begin to cross the reverse shock.  Here we present new optical and infrared (IR) spectra of SN 2012ca, along with a number of other SNe~Ia-CSM, SNe~IIn, and SNe~Ia.  We compare the line identifications of \citet{inserra14} among the various spectra in our database.  In particular, we consider the case of the SN~Ia-CSM 2013dn, for which we have well-sampled, high-S/N spectra through day $\sim$441 post-maximum.  Section \ref{sec_obs} presents the observations, while \S \ref{sec_analysis} offers a detailed comparison of the spectra to other SN types.  In \S \ref{sec_discussion} we discuss the implications on the SN~Ia-CSM progenitor, and \S \ref{sec_conclusion} summarises our conclusions.

\begin{table}
\centering
\caption{Summary of Distances to {\it Reported} SNe~Ia-CSM \label{tab_100mpc}}
\begin{tabular}{ l c c }
\hline
SN & Distance &  Reference\\
      &  (Mpc)      &\\
\hline
2014ab & 95 & ATel 4076\\
2014T & 375 & CBET 3815\\
2014Y & 162 & CBET 3824\\ 
2013dn & 233 & CBET 3570\\ 
2013I & 144 & CBET 3386\\
2012ca & 79 & CBET 3101\\
CSS120327:110520-015205 & 215 & ATel 4081\\
2011dz & 100 & CBET 2761\\
2011jb & 142 & CBET 2947\\
PTF11kx & 194 & Dilday et al. 2012\\
PTF10htz & 147 & Silverman et al. 2013b\\
2009in & 98 & CBET 1953\\
2008J & 92 & CBET 1218\\
\hline
\end{tabular}
\end{table}

\section{Observations}
\label{sec_obs}

\begin{table*}
\centering
\caption{RATIR Photometry of SN 2013dn \label{tab_ratir_phot}}
\begin{tabular}{ l c c c c c c c c }
\hline
JD $-$ & Epoch &  $r$ &  $i$ &  $Z$ &  $Y$ &  $J$ & $H$ & Log \\
2,450,000 &  (days) & \multicolumn{6}{c}{mag} & $L_{\rm opt}$/(erg s$^{-1}$) \\ 
\hline
6510 & 106  & 17.00 (0.02) & 17.01 (0.02) & 16.61 (0.02) & 16.78 (0.03) & 17.16 (0.03) & 17.18 (0.04) & 43.12 \\
6519 & 115  & 17.25 (0.02) & 17.01 (0.02) & 16.74 (0.02) & 16.92 (0.03) & 17.37 (0.03) & 17.26 (0.04) & 43.39 \\
6539 & 135  & 17.30 (0.02) & 17.20 (0.02) & 16.93 (0.02) & 17.08 (0.03) & 17.56 (0.06) & 17.44 (0.08) & 43.29 \\
6546 & 142  & 17.48 (0.02) & 17.40 (0.02) & 17.06 (0.02) & 17.21 (0.03) & --                    & -- & 43.26 \\
6551 & 147  & 17.42 (0.02) & 17.34 (0.02) & 16.96 (0.02) & 17.25 (0.03) & 17.77 (0.05) & 17.59 (0.08) &  43.23 \\
6562 & 158  & 17.53 (0.02) & 17.43 (0.02) & 17.03 (0.02) & 17.29 (0.03) & 17.86 (0.05) & 17.75 (0.07) &  43.20 \\
6579 & 175  & 17.85 (0.02) & 17.61 (0.02) & 17.15 (0.03) & 17.54 (0.04) & 18.13 (0.05) & 17.98 (0.08) &  43.11\\
6591 & 187  & 18.04 (0.02) & 17.73 (0.02) & 17.21 (0.02) & 17.72 (0.04) & 18.10 (0.05) & 18.11 (0.08) & 43.09\\
6606 & 202  & 17.99 (0.02) & 17.88 (0.02) & 17.34 (0.02) & 17.98 (0.04) & 18.35 (0.06) & 18.44 (0.09) & 43.02 \\
6612 & 208  & 18.01 (0.02) & 17.90 (0.02) & 17.93 (0.03) & 18.09 ( 0.04) & 18.30 (0.06) & 18.21 (0.09) & 42.98 \\
6621 & 217 &  18.11 (0.02) & 18.03 (0.02) & 17.47 (0.03) & 18.18 (0.04) & 18.44 (0.06) & 18.72 (0.09) & 42.95 \\
6632 & 228 & 18.16 (0.02) &  18.08 (0.02) & 17.51 (0.03) & 18.31 (0.04) & 18.48 (0.06) & 18.77 (0.09) & 42.94 \\
6638 & 234 & 18.30 (0.02) &  18.21 (0.02) & 17.62 (0.03) & 18.27 (0.04) & 18.60 (0.06) & 18.71 (0.09) & 42.89 \\
6680 & 276 & 18.57 (0.02) &  18.46 (0.02) & 17.90 (0.03) & 18.59 (0.04) & 18.07 (0.09)  & 18.13 (0.19) &  42.81\\
6821 & 417 & 19.95  (0.03) & 19.97 (0.04) & 19.49 (0.05) & --                    & $>$19.00       & $>$18.75      & 42.26 \\
\hline
\end{tabular}
\end{table*}

This paper presents new data on SNe~Ia-CSM 2012ca and 2013dn and SNe~IIn 2005ip and 2009ip.  SN 2012ca was discovered in the late-type spiral galaxy ESO 336-G009 on 2012 Apr.~25.6 (UT dates are used throughout this paper) at redshift $z=0.019$~\citep[$m_r \approx 14.8$ mag;][]{drescher12,inserra14}.  The earliest spectrum matches that of SN~Ia-CSM 1997cy at an estimated $\sim 60$ days post-maximum \citep{inserra12, valenti12}, although the peak of SN 1997cy was never observed \citep{germany00,turatto00}.  Similar to \citet{inserra14}, we take the light-curve peak to be 2012 Mar.~2.

SN 2013dn was discovered in the galaxy PGC 71942 on 2013 Jun.~14.45 at $z=0.056$~\citep[$m_r \approx 16.2$ mag;][]{drake13}.  The earliest spectrum (2013 Jun.~26.13) matches that of SN~Ia-CSM 2005gj at 54 days post-explosion \citep{drake13}, although at this redshift the derived absolute magnitude ($\sim -21.1$) is somewhat brighter than that derived for SN 2005gj around maximum light.  Given that SN 2005gj peaks $\sim 32$ days post-explosion in the $r$ \citep{prieto07}, we take 2013 Jun.~4 to be the $r$-band light-curve peak for SN 2013dn.

SN 2005ip was discovered in NGC 2906 on 2005 Nov.~5 at $z=0.0072$~\citep{boles05}, although early-time optical spectra suggested the discovery occurred a few weeks following the explosion \citep{modjaz05}.  The spectra, dominated by narrow hydrogen emission lines, led to a Type IIn classification \citep{fox09,smith09ip}.  While there are no precise constraints on the explosion date, \citet{smith09ip} and \citet{fox09} suggest the peak may have been $\sim 2$ weeks prior to the discovery.  We therefore take 2005 Oct.~22 for the light-curve peak.

SN 2009ip was originally classified as a SN \citep{maza09}, but was actually discovered during a luminous blue variable star (LBV) outburst and did not undergo its most extreme outburst until 2012 August \citep{mauerhan13,fraser13a,fraser13b,pastorello13,prieto13,smith13ip,smith14ip,graham14,levesque14,margutti14}.  The progenitor has not been confirmed to have disappeared, but evidence suggests the star underwent a terminal explosion \citep{mauerhan14,smith14ip}. The progenitor was probably a massive star (i.e., not a white dwarf), given (1) high and continuous pre-SN wind speeds ($> 600$ km s$^{-1}$), (2) a luminous ($10^6$~\lsolar) and eruptive progenitor \citep{smith10ip}, and (3) the presence of a hot LBV spectrum during the earlier ``impostor'' phases \citep{smith10ip,foley11,pastorello13}.  Similar to \citet{smith14ip}, we take the light-curve peak to be 2012 Sep.~24.

\begin{figure}
\centering
\includegraphics[width=3.3in]{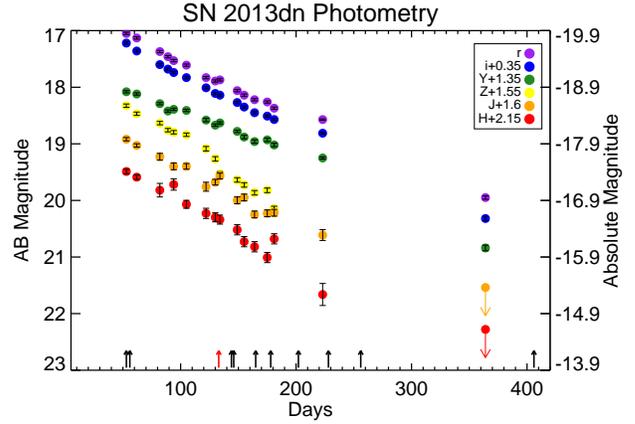}
\caption{Optical/IR light curve of SN 2013dn.  Black arrows signify epochs at which optical spectra were obtained.  The red arrow marks the epoch at which the near-IR spectrum was obtained.}
\label{fig_ratir_phot} 
\end{figure}

All epochs presented in this article correspond to days post-maximum unless otherwise specified.

\subsection{Optical and Near-Infrared Photometry}

Photometry of SN 2013dn was obtained with the multi-channel Reionization And Transients InfraRed camera \citep[RATIR;][]{butler12} mounted on the 1.5-m Johnson telescope at the Mexican Observatorio Astrono\'mico Nacional on Sierra San Pedro M\'artir in Baja California, M\'exico \citep{watson12}. Typical observations include a series of 80-s exposures in the $ri$~bands and 60-s exposures in the $ZYJH$~bands, with dithering between exposures.  RATIR's fixed IR filters cover half of their respective detectors, automatically providing off-target IR sky exposures while the target is observed in the neighbouring filter. Master IR sky frames are created from a median stack of off-target images in each IR filter.  No off-target sky frames were obtained on the optical CCDs, but the small galaxy size and sufficient dithering allowed for a sky frame to be created from a median stack of all the images in each filter.  Flat-field frames consist of evening sky exposures. Given the lack of a cold shutter in RATIR's design, IR dark frames are not available.  Laboratory testing, however, confirms that the dark current is negligible in both IR detectors \citep{fox12}.

\begin{figure}
\centering
\includegraphics[width=3.3in]{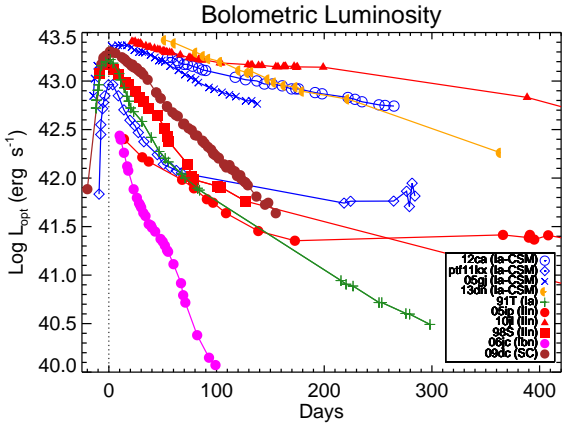}
\caption{Bolometric light curve of SN 2013dn (orange) compared to that of several well-known SNe~IIn (red), SNe~Ia-CSM (blue), and the overluminous Type Ia SN 1991T (green).  Although both the SNe~Ia-CSM and SNe~IIn have late-time light-curve plateaus caused by ongoing CSI, the peak and plateau luminosities can vary significantly within each class.  The total radiated luminosity output for SN 2013dn (and several other SNe~Ia-CSM) is similar to that of some of the SNe~IIn ($\sim$ few $\times 10^{50}$~erg), which is still consistent with a thermonuclear explosion of a white dwarf ($\sim10^{51}$~erg) but requires a high conversion efficiency ($\sim 0.5$).
}
\label{fig_lightcurve} 
\end{figure}

The data were reduced, coadded, and analysed using standard CCD and IR processing techniques in IDL and Python, utilising online astrometry programs {\tt SExtractor} and {\tt SWarp}\footnote{SExtractor and SWarp can be accessed from http://www.astromatic.net/software.}. Calibration was performed using field stars with reported fluxes in both 2MASS \citep{skrutskie06} and the Sloan Digital Sky Survey (SDSS) Data Release 9 Catalogue \citep{ahn12}.  Figure \ref{fig_ratir_phot} plots (and Table \ref{tab_ratir_phot} lists) the optical and near-IR photometry obtained for SN 2013dn.  For a comparison, Figure \ref{fig_lightcurve} plots the bolometric luminosity against that of other relevant SNe we will be comparing.

\subsection{Optical Spectroscopy}

\begin{figure}
\centering
\vspace{-0.15in}
\includegraphics[width=3.6in]{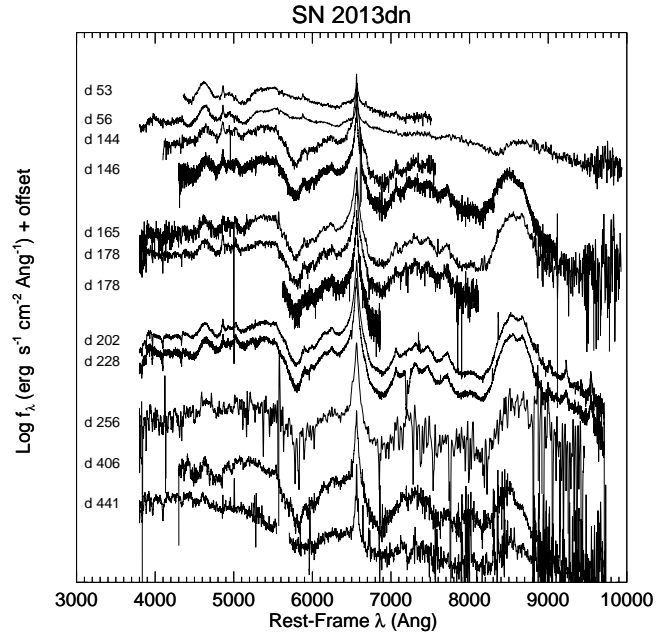}
\caption{Time series of optical spectra of the Type Ia-CSM SN 2013dn.
}
\label{fig_13dn_all} 
\end{figure}

Table \ref{tab_opt_spectra} summarises the details concerning new optical spectra presented in this paper.  Figure \ref{fig_13dn_all} plots all the spectra of SNe 2013dn.  Some data were obtained with the Kast double spectrograph on the Shane 3-m telescope at Lick Observatory \citep{miller93}, the dual-arm Low Resolution Imaging Spectrometer \citep[LRIS;][]{oke95} mounted on the 10-m Keck~I telescope, and the DEep Imaging Multi-Object Spectrograph \citep[DEIMOS;][]{faber03} on the 10-m Keck~II telescope.  Most observations had the slit aligned along the parallactic angle to minimise differential light losses \citep{filippenko82}.  

Additional spectra were obtained with the Hiltner 2.4~m telescope at MDM Observatory using the Boller \& Chivens CCDS spectrograph.\footnote{http://www.astronomy.ohio-state.edu/MDM/CCDS/.} The 150 line mm$^{-1}$ grating was used with a $1.2''$ slit to yield spectra having a full width at half-maximum intensity (FWHM) resolution of 12~\AA.  Data were also obtained from the MMT 6.5-m telescope using the Blue Channel (BC) spectrograph \citep{schmidt89}. The 300 and 1200 line mm$^{-1}$ gratings were used in conjunction with a $1.0''$ slit to yield spectra having FWHM resolutions of 7 and 2~\AA, respectively.

\begin{table}
\centering
\caption{Summary of Optical Spectra \label{tab_opt_spectra}}
\begin{tabular}{ l c c c c}
\hline
SN & JD $-$ & Epoch & Instrument & Int\\
      & 2,450,000 & (days) &  & (s) \\
\hline
2013dn &    6479    &     53  &  MDM/MkIII    &   2700  \\   
& 6482 & 56 & Lick/Kast & 900 \\
&    6570    &    144  &  MDM/CCDS     &   3600  \\
& 6572 & 146 & Keck/DEIMOS & 2400\\
& 6591 & 165 & Lick/Kast & 2400\\
&    6604    &    178  &  MMT/BC-300   &   1200  \\ 
&    6604    &    178  &  MMT/BC-1200  &   2400  \\
& 6628 & 202 & Keck/LRIS & 2400\\
& 6654 & 228 & Keck/DEIMOS & 2400\\
& 6682 & 256 & Lick/Kast & 2400\\
& 6834 & 406 & Keck/DEIMOS & 1200\\
& 6834 & 441 & Keck/LRIS & 1200\\
\hline
2012ca & 6484 & 502 & Keck/DEIMOS & 600 \\
& 6506 & 524 & Keck/DEIMOS & 600 \\
\hline
2009ip & 6834 & 434 & Keck/DEIMOS & 1800\\ 
\hline
\end{tabular}
\end{table}

The spectra were reduced using standard techniques \citep[e.g.,][]{foley03,silverman12bsnip1}. Routine CCD processing and spectrum extraction were completed with {\tt IRAF}\footnote{IRAF: the Image Reduction and Analysis Facility is distributed by the National Optical Astronomy Observatory, which is operated by the Association of Universities for Research in Astronomy (AURA), Inc., under cooperative agreement with the US National Science Foundation (NSF).}, and the data were extracted with the optimal algorithm of \citet{horne86}. We obtained the wavelength scale from low-order polynomial fits to calibration-lamp spectra. Small wavelength shifts were then applied to the data after cross-correlating a template-sky spectrum to an extracted night-sky spectrum. Using our own IDL routines, we fit a spectrophotometric standard-star spectrum to the data in order to flux calibrate the SN and to remove telluric absorption lines \citep{wade88,matheson00}.

Other optical spectra used throughout this paper were obtained from both the Berkeley Supernova Database \citep{silverman12bsnip1} and the Weizmann Interactive Supernova data REPository \citep[WISeRep;][]{yaron12}.

\subsection{Near-Infrared Spectroscopy}

Table \ref{tab_ir_spectra} summarises the details concerning new near-IR spectra presented in this paper.  Some data were obtained with with TripleSpec spectrographs \citep{wilson04,herter08} operating at both the Apache Point Observatory 3-m and the Palomar Observatory 5-m telescopes.  TripleSpec observations typically consisted of 300-s exposures, taken at varying locations along the slit and then pair-subtracted to allow for the correction of night-sky emission lines.  We extract the spectra with a modified version of the IDL-based {\it SpexTool} \citep{cushing04}.  The underlying galaxy and sky emission are approximated by a polynomial fit and subtracted from the supernova.  A-type calibration stars were observed directly after each SN exposure to remove telluric absorption lines using the IDL-based {\it xtellcor} package \citep{vacca03}.  The day $+$917 of SN 2010jl was previously published by \citet{borish14}, along with the full near-IR evolution of SN 2005ip from earlier epochs.

\begin{table}
\centering
\caption{Summary of Infrared Spectra \label{tab_ir_spectra}}
\begin{tabular}{ l c c c c c c}
\hline
SN & JD $-$ & Epoch & Instrument \\
& 2,450,000 & (days) &  \\
\hline
2014J &6727 & 36 &TripleSpec (APO) \\ 
& 6786 & 95 &TripleSpec (APO) \\ 
\hline
2013dn &6559 & 133 &TripleSpec (APO) \\ 
\hline
2012ca & 6053 & 70 & FIRE \\ 
& 6087 & 104 & FIRE \\ 
\hline
2010jl & 5657 & 170 &TripleSpec (APO)\\ 
& 6727 & 1233 &TripleSpec (APO) \\ 
\hline
2005ip & 4547 & 884 &TripleSpec (APO) \\ 
& 4580 & 917 &TripleSpec (APO) \\ 
& 6727 & 3064 &TripleSpec (APO) \\ 
\hline
2008J &4714 & 234 &TripleSpec (APO) \\ 
\hline
2007gr & 4376 & 35 &TripleSpec (Palomar) \\ 
& 4450 & 109 &FIRE \\ 
\hline
\end{tabular}
\end{table}

Other IR spectra were obtained with the Folded-port InfraRed Echellette (FIRE) spectrograph at the Magellan 6.5-m telescope \citep{simcoe08}.  The FIRE spectra were obtained in the high-throughput prism mode with a 0$\farcs$6 slit.  This configuration yields a continuous wavelength
coverage from 0.8 to 2.5 \micron~with a resolution of $R \approx 500$ in the $J$ band.  When acquiring the supernova, the slit was oriented at the parallactic angle to minimise the effect of atmospheric dispersion
\citep{filippenko82}.  At each epoch, several frames were obtained using the conventional ABBA ``nod-along-the-slit'' technique and the ``sampling-up-the-ramp'' readout mode.  The per-frame exposure time was between 95.1 and 158.5~s, depending on the brightness of the supernova.  These exposure times were chosen such that adequate signal was obtained in each frame without saturating the bright night-sky lines in the $K$ band.  At each epoch, an A0V star was observed close to the science observations in time, angular distance, and airmass for telluric correction, as per the method described by \citet{vacca03}.

The data were reduced using the IDL pipeline \texttt{firehose}, specifically designed for the reduction of FIRE data.  The pipeline performed steps of flat fielding, wavelength calibration, sky subtraction, spectral tracing, and spectral extraction.  The sky flux was modeled using off-source pixels as described by \citet{kelson03} and subtracted from each frame. Next, the spectral extraction was performed using the optimal technique \citep{horne86}, a weighting scheme that delivers the maximum S/N while preserving spectrophotometric accuracy.  Individual spectra were then combined with sigma clipping to reject spurious pixels.  Corrections for telluric absorption were performed using the IDL tool \texttt{xtellcor} developed by \citet{vacca03}.  To construct a telluric correction spectrum free of stellar absorption features, a model spectrum of Vega was used to match and remove the hydrogen lines of the Paschen and Brackett series from the A0V telluric standard.  The resulting telluric correction spectrum was also used for flux calibration.

\begin{figure}
\centering
\includegraphics[width=3.3in]{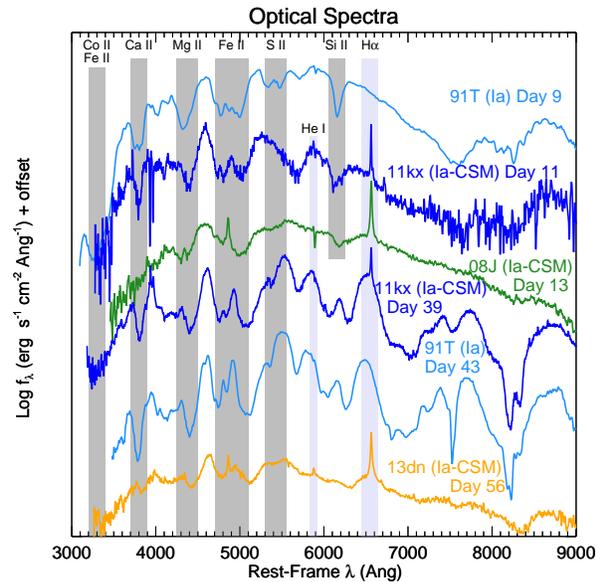}
\caption{Comparison of early-time spectra, including the Type Ia SN 1991T (light blue), the Type Ia-CSM PTF11kx (dark blue), the Type IIn/Ia-CSM SN 2008J (green), and the Type Ia-CSM SN 2013dn (orange).  Vertical grey bars highlight the position of several broad absorption lines associated with SNe~Ia, while vertical light-blue bars highlight the position of narrow H and He emission associated with SNe~IIn.  SN 2013dn looks very similar to SN 2008J, which was decomposed into a combination of a SN 1991T-like event and a blackbody continuum \citep{taddia12}.
}
\label{fig_opt_early} 
\vspace{-0.2in}
\end{figure}

\begin{figure*}
\centering
\includegraphics[width=3.2in]{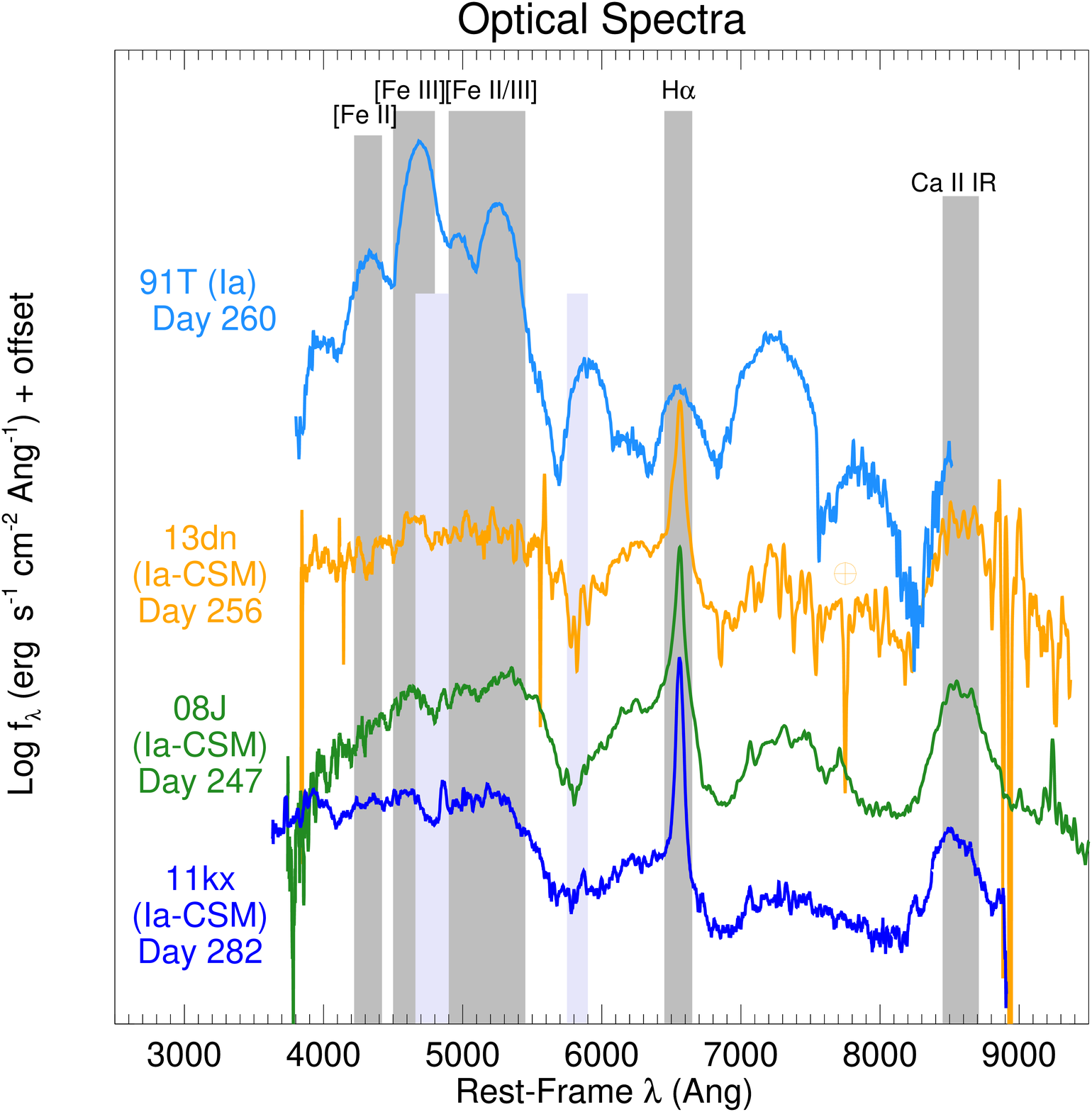}
\includegraphics[width=3.2in]{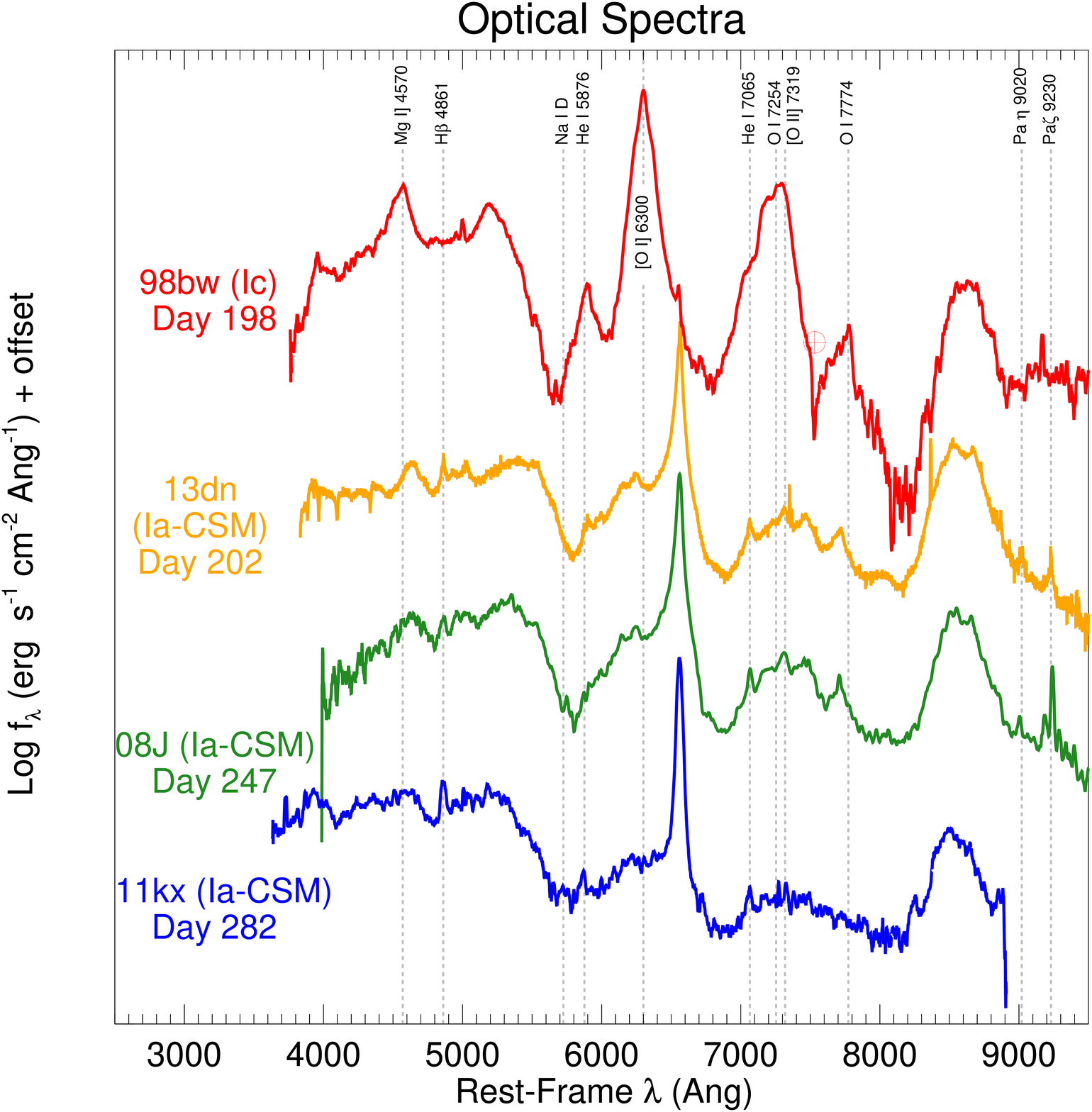}
\caption{(left) Comparison of late-time spectra, including the Type Ia SN 1991T (light blue), the Type Ia-CSM PTF11kx (dark blue), the Type IIn/Ia-CSM SN 2008J (green), the Type Ia-CSM SN 2013dn (orange).  (right)  Comparison of the interacting SNe~Ia-CSM to the broad-lined Type Ic SN 1998bw (red).  Dashed vertical grey lines highlight several of the more prominent narrow features, solid grey vertical bars mark a few of the broader features, and solid blue vertical bars show some of the more prominent P Cygni features.  Not all lines appear in all spectra.  Also note that the 6500 \AA\ feature in SN 1991T is a blend of [\ion{Fe}{II}] and Co, not H$\alpha$.
}
\label{fig_opt_late} 
\end{figure*}

\section{Analysis of the Spectra}
\label{sec_analysis}

\subsection{Early-Time Optical Spectra}
\label{sec_optev}

Figure \ref{fig_opt_early} illustrates the early-time evolution of several SNe ranging from days $+$9 to 56, including the SNe~Ia-CSM 2013dn (orange), 2008J (green), and PTF11kx (dark blue), and the overluminous SN~Ia 1991T (light blue).  Owing to difficulties in identifying and removing the continuum, we instead apply a minimal degree of artificial reddening or dereddening for plotting purposes in Figure \ref{fig_opt_early}.  This reddening/dereddening does not change the presence of any spectral lines and is not applied during the analysis of the spectra elsewhere in the article (although we deredden SN 2008J throughout the paper by a factor consistent with measurements of \citet{taddia12}).  Vertical grey bars highlight the position of several broad absorption lines associated with SNe~Ia, while vertical light blue bars mark the position of narrow H and He emission associated with SNe~IIn.   At these early times,  the overall blue colour of the spectra is caused by the continuum from the underlying photosphere.

SN 1991T (day $+$9) exhibits the deepest absorption lines, but this spectrum was also obtained closest to peak brightness.  Even so, SN 1991T-like objects tend to have weaker lines than do normal SNe~Ia or underluminous SNe~Ia (e.g., SN 1991bg; \citealt{silverman12bsnip2}).  The corresponding absorption lines are still noticeable in SNe~Ia-CSM, particularly PTF11kx, despite being weakened by the underlying continuum and H/He emission lines.  Earlier spectra of SN 2008J have even more pronounced absorption features and can be decomposed into a combination of a SN 1991T-like event and a blackbody continuum \citep{taddia12}.  The earliest spectrum we obtained of SN 2013dn was on day $+$56, but even at this epoch the spectrum shows remarkable similarity to SN 2008J on day $+$13.  Furthermore, despite the varying degrees of underlying continuum and the S/N, these two SNe share many of the same features as PTF11kx.  

\subsection{Late-Time Optical Spectra}

\begin{figure*}
\centering
\includegraphics[width=7in]{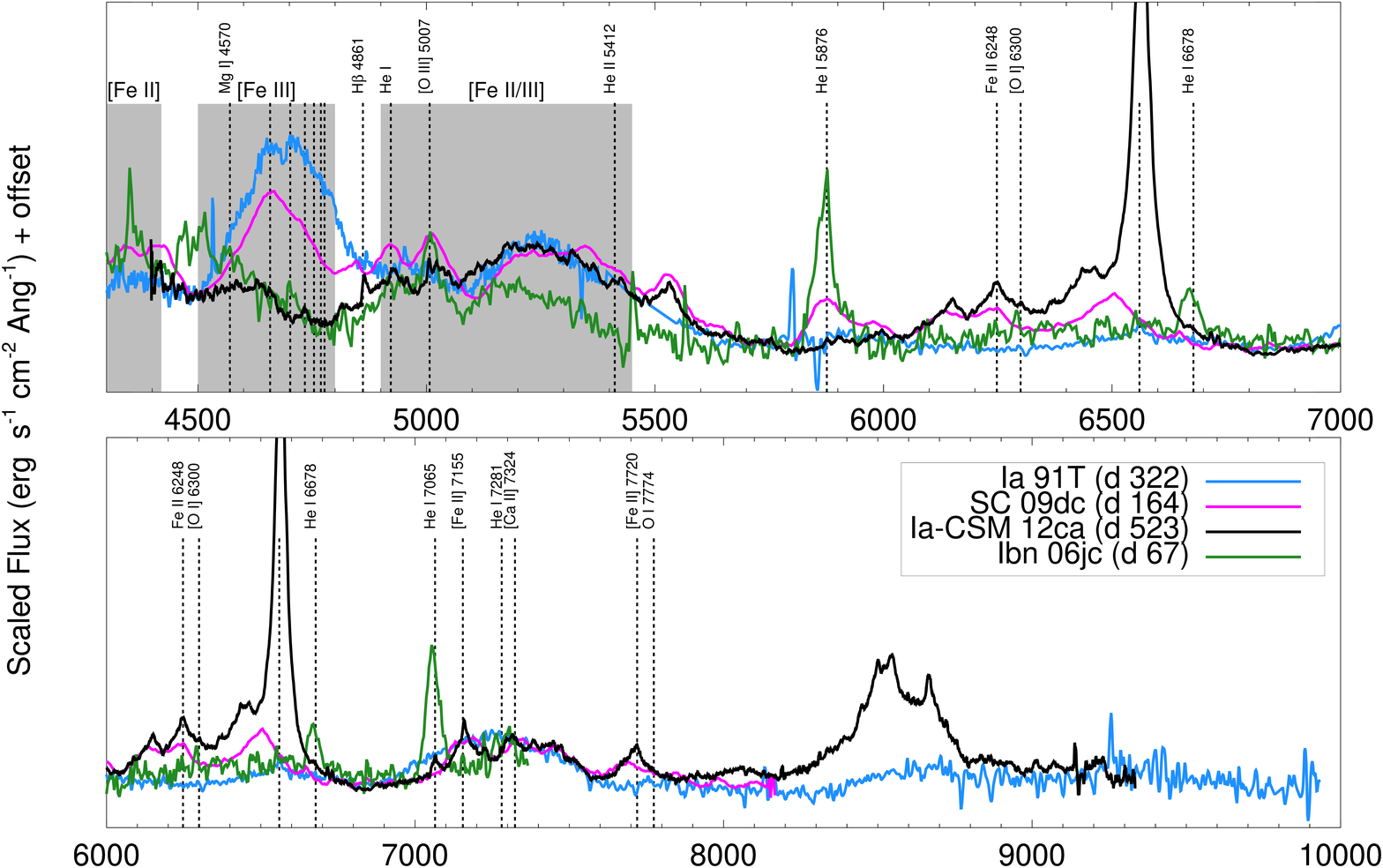}\\
\includegraphics[width=7in]{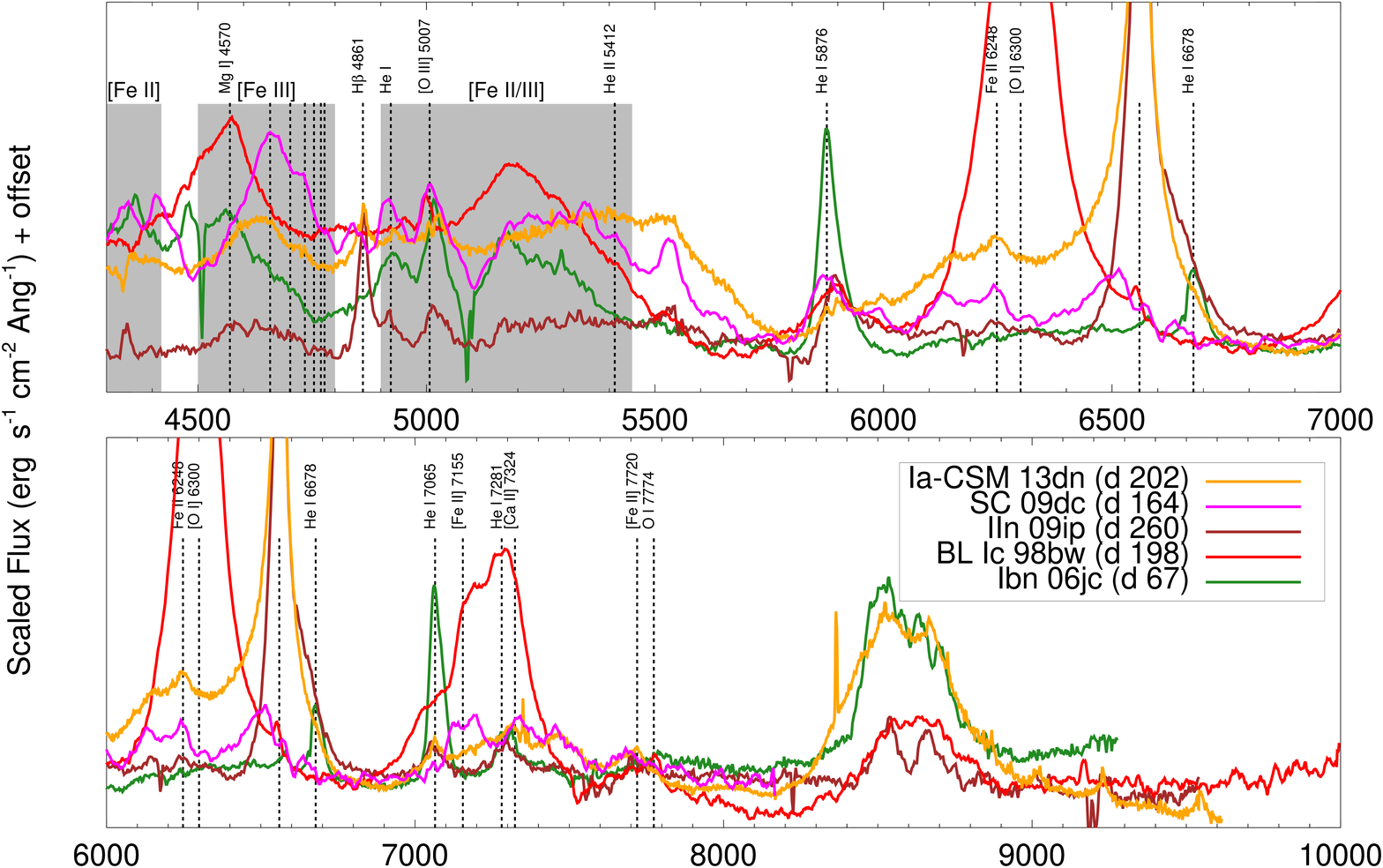}
\vspace{0.2in}
\caption{(top) Comparison of SN 2012ca (black) at later epochs to the Type Ia SN 199T (light blue), the super-Chandrasekhar-mass (CS) candidate Type Ia SN 2009dc (pink), and the Type Ibn SN 2006jc (green).  (bottom) Comparison of SN 2013dn (orange) at $\sim 200$ days to the CS candidate Type Ia SN 2009dc (pink), the Type IIn SN 2009ip (brown), the Type Ibn SN 2006jc (green), and the broad-lined Type Ic SN 1998bw (red).  Dashed vertical lines highlight several of the more prominent narrow features, while solid grey vertical bars mark a few of the broader iron features.  Unlabeled dotted black lines correspond to [\ion{Fe}{III}].
}
\label{fig_opt_late_lin_300} 
\end{figure*}

Figure \ref{fig_opt_late} compares the SNe~Ia-CSM at later epochs, ranging from days $+$202 to 282.  We also consider the Type Ia SN 1991T (light blue) and the broad-lined Type Ic SN 1998bw (red) given their widespread use in other papers \citep[e.g.,][]{inserra14}.  The reader should be aware, however, of several conditions that make a direct comparison difficult.  First, strong circumstellar lines and an underlying continuum dominate the spectra of the interacting SNe, but are not necessarily associated with the exploding star or ejecta.  A cold, dense shell (CDS) can form between the forward and reverse shocks that, if optically thick, may further obscure the underlying ejecta.  Second, the continuum levels are difficult to identify since the blue flux shortward of 5500~\AA\ in the SNe~Ia-CSM may arise from a ``quasi-continuum'' produced by many overlapping permitted and forbidden lines of iron-group elements \citep[e.g.,][]{deng04,bowers97,branch08,silverman13b}.  Third, the nebular spectra of SNe~Ia do not have many unique or unambiguous characteristic features among various SN~Ia subtypes (see Fig.~13 of \citealt{parrent14}).  We discuss the implications of these effects on our interpretation in \S \ref{sec_discussion}.

Figure \ref{fig_opt_late_lin_300} compares different SN types on a linear scale over different epochs at late-times.  The linear scale highlights the relative line intensities more clearly than the log scale in Figure \ref{fig_opt_late}.  All spectra are scaled by a multiplicative factor to highlight features relative to the likely continuum, which we identify as being just redward of H$\alpha$ ($\sim$6800 \AA) and just blueward of He I $\lambda$5876.  The SNe~Ia-CSM have three distinguishing traits: (1) a low [\ion{Fe}{III}] $\lambda$4700/[\ion{Fe}{II}] $\lambda$5200 ratio, (2) weak [\ion{O}{I}] $\lambda$6300, and (3) weak \ion{Mg}{I}] $\lambda$4570.  The lack of obvious iron (thermonuclear) and/or intermediate mass (core-collapse) signatures is the very characteristic that led to the ambiguity surrounding the SNe~Ia-CSM in the first place.

\begin{figure}
\centering
\includegraphics[width=3.3in]{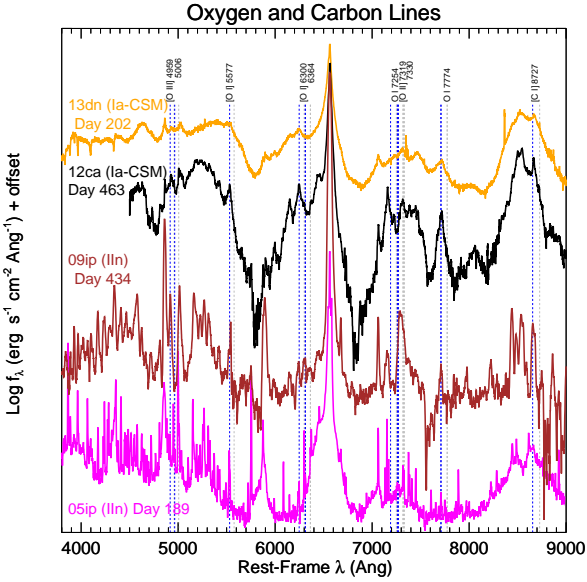}
\caption{Comparison of late-time spectra, including the Type Ia-CSM SN 2013dn (orange), the Type Ia-CSM SN 2012ca (black), the Type IIn SN 2009ip (brown), and the Type IIn SN 2005ip (pink).  \citet{inserra14} identify oxygen and carbon in the spectrum of SN 2012ca, all blueshifted by $\sim 2500$ \kms.  Dashed vertical lines highlight the positions of these spectral lines at rest (grey) and blueshifted (dark blue).  Surprisingly, these spectral lines also appear in SNe 2005ip, 2013dn, and 2008J at the same blueshift.  Unlikely to be a coincidence, in the text we argue alternative line identifications.
}
\label{fig_inserra} 
\end{figure}

\subsubsection{Low [\ion{Fe}{III}]/[\ion{Fe}{II}] ratio}

\citet{inserra14} point out, in favour of their core-collapse progenitor argument, that the dominant [\ion{Fe}{III}] features observed in most nebular SNe~Ia (e.g., SN 1991T and even PTF11kx) are not observed in SN 2012ca, even after accounting for dilution by a continuum.  Figure \ref{fig_opt_late_lin_300} highlights blended [\ion{Fe}{III}] lines at $\sim$4700 \AA, which \citet{fesen96} decompose into a number of individual components at 4658.10, 4701.62, 4733.93, 4754.83, 4769.60, and 4777.88 \AA.   Another iron blend, composed primarily of [\ion{Fe}{II}], is centered at $\sim 5200$ \AA.  Indeed, the [\ion{Fe}{III}]/[\ion{Fe}{II}] ratio in SNe 2012ca and 2013dn is significantly lower than that in SN~Ia 1991T.  The ratio is similarly low, however, in the SC candidate SN 2009dc, the SN~Ibn 2006jc, and the SN~IIn 2009ip, although the strengths of the features do vary.  We further discuss these comparisons in \S \ref{sec_discussion}.

\begin{figure}
\centering
\vspace{0.25in}
\includegraphics[width=2.7in]{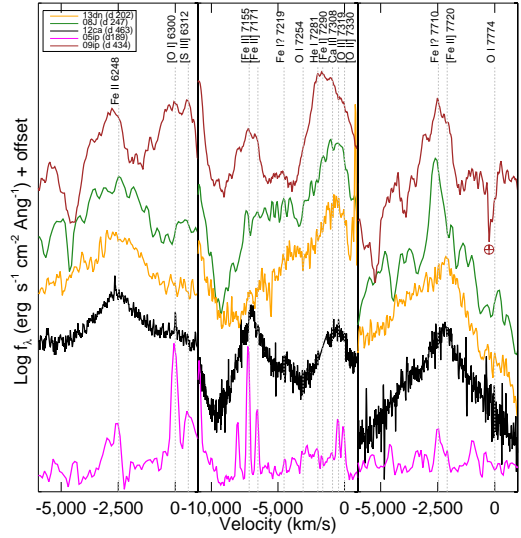}
\vspace{0.1in}
\caption{Velocity profiles of spectra plotted in Figure \ref{fig_inserra}, plus the Type IIn/Ia-CSM SN 2008J (green), relative to the rest-frame [\ion{O}{I}] $\lambda$6300, [\ion{O}{II}] 7330, and \ion{O}{I} 7774, respectively.  Vertical grey dashed lines mark the primary oxygen lines at rest velocity.  At a blueshift of $\sim 2500$ \kms, several other line identifications are possible.  In fact, these lines are all identified in the Type IIn SN 2005ip \citep{smith09ip}.  Only SN 2005ip shows any evidence for oxygen, and in this case only a relatively narrow line at 6300 \AA.
\label{fig_oxy} }
\end{figure}


\subsubsection{No Oxygen or Carbon Detected}
\label{sec_oxy}

\citet{inserra14} further argue for a core-collapse origin of SN 2012ca based on the identification of (1) oxygen, (2) carbon, (3) magnesium, and (4) helium.  We examine the oxygen and carbon lines here in more detail, and discuss magnesium and helium below in \S \ref{sec_ir_spectra}.  Figure \ref{fig_inserra} plots SN 2012ca (black) alongside the SN~Ia-CSM 2013dn (orange) and SNe IIn 2005ip (pink) and 2009ip (brown).  Note that the late-time spectra of SNe 2009ip and 2012ca ($> 400$ days) likely reflect the nebular phase during which the emission lines are generated primarily by the ejecta.  In this case, we believe the emission lines are less dominated by CSI and more representative of the SN ejecta.

{\it Oxygen:} \citet{inserra14} specifically identify the \ion{O}{I} $\lambda$7774, [\ion{O}{I}] $\lambda\lambda$6300, 6364, and [\ion{O}{II}] $\lambda$7254 lines, all blueshifted by $\sim 2500$ \kms.  Blueshifted lines can be expected in late-time spectra as absorption from the dense ejecta obscures emission from the receding SN hemisphere.  \citet{milisavljevic12} observe this effect in a number of late-time core-collapse SN spectra, but the blueshifted velocities range from 500 \kms~up to $> 3000$ \kms.  \citet{silverman13b} point out that these lines are indeed lacking from their SN~Ia-CSM sample (although they do concede that part of the very broad emission feature around 7400~\AA\ may be a blend of [\ion{O}{II}] $\lambda\lambda$7319, 7330 and [\ion{Ca}{II}] $\lambda\lambda$7291, 7324). 

\begin{figure}
\centering
\begin{center}
\includegraphics[width=3in]{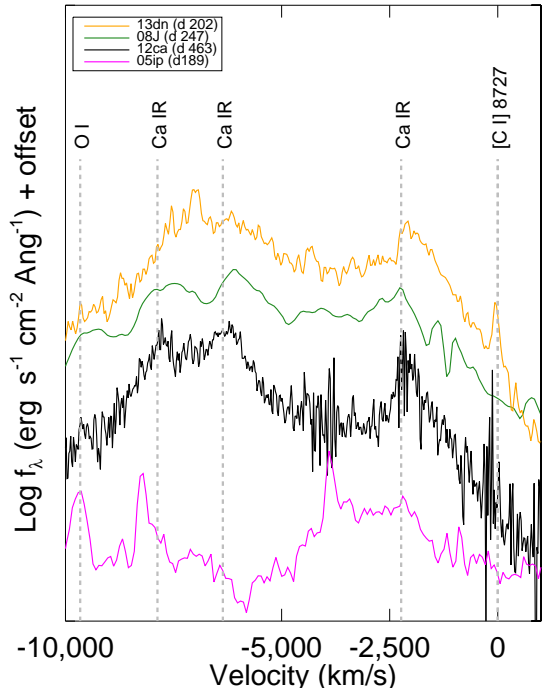}
\caption{Velocity profiles of spectra plotted in Figure \ref{fig_inserra}, plus the Type IIn/Ia-CSM SN 2008J (green).  Vertical grey dashed lines mark the primary carbon line at rest velocity.  At a blueshift of $\sim 2500$ \kms, a line consistent with \ion{Ca}{II} exists.  This line, in fact, is the dominant line identified in the spectra presented by \citet{inserra12}.  Some carbon does exists, however, in SNe 2013dn and 2012ca, but not in the Type IIn SN 2005ip.
}
\label{fig_cair} 
\end{center}
\end{figure}

Vertical dashed lines in Figure \ref{fig_inserra} mark these oxygen lines both in the rest frame (grey) and blueshifted (blue), while Figure \ref{fig_oxy} plots the corresponding velocity profiles.  Taken on its own, SN 2012ca does exhibit strong lines that could be considered consistent with blueshifted oxygen, but we identify the $\lambda$7774 and $\lambda\lambda$6300, 6364 lines at the exact same blueshift in SNe 2005ip, 2013dn, 2008J, 2009ip, and 1998S.  The lack of any velocity spread between the SNe is curious, especially considering that the [\ion{O}{II}] $\lambda$7254 line is missing in SNe 2013dn and 2008J.  We suggest the lines that appear to be consistent with blueshifted oxygen actually correspond to broad iron lines that we describe in \S \ref{sec_ionization} below.  Nonetheless, we note that all of the spectra do show evidence for weaker and unshifted O~I $\lambda$7774, [O~I] $\lambda$6300, and O~I $\lambda$8446 with FWHM $\approx 1000$ \kms.

{\it Carbon:} \citet{inserra14} also identify the [C~I] $\lambda$8727 line after day $+$256.  Figure \ref{fig_inserra} again marks this line both in the rest frame (grey) and blueshifted (blue).  Figure \ref{fig_cair} plots the velocity profiles for SNe 2005ip, 2012ca, 2008J, and 2013dn.  If the carbon were expected to be found in the ejecta, like the oxygen, then we would should see a $-2500$ \kms\ shift.  Indeed, there is an emission feature at $-2500$ \kms, but this line is consistent with the Ca~IR triplet.  We do detect emission at $\sim 0$ \kms\ in SNe 2012ca and 2013dn, but this line is (1) at 0 \kms\ and (2) narrow, both of which would be unexpected if associated with the blueshifted ejecta. 

\subsubsection{High-Ionisation Coronal Lines}
\label{sec_ionization}

\begin{figure*}
\centering
\includegraphics[width=7in]{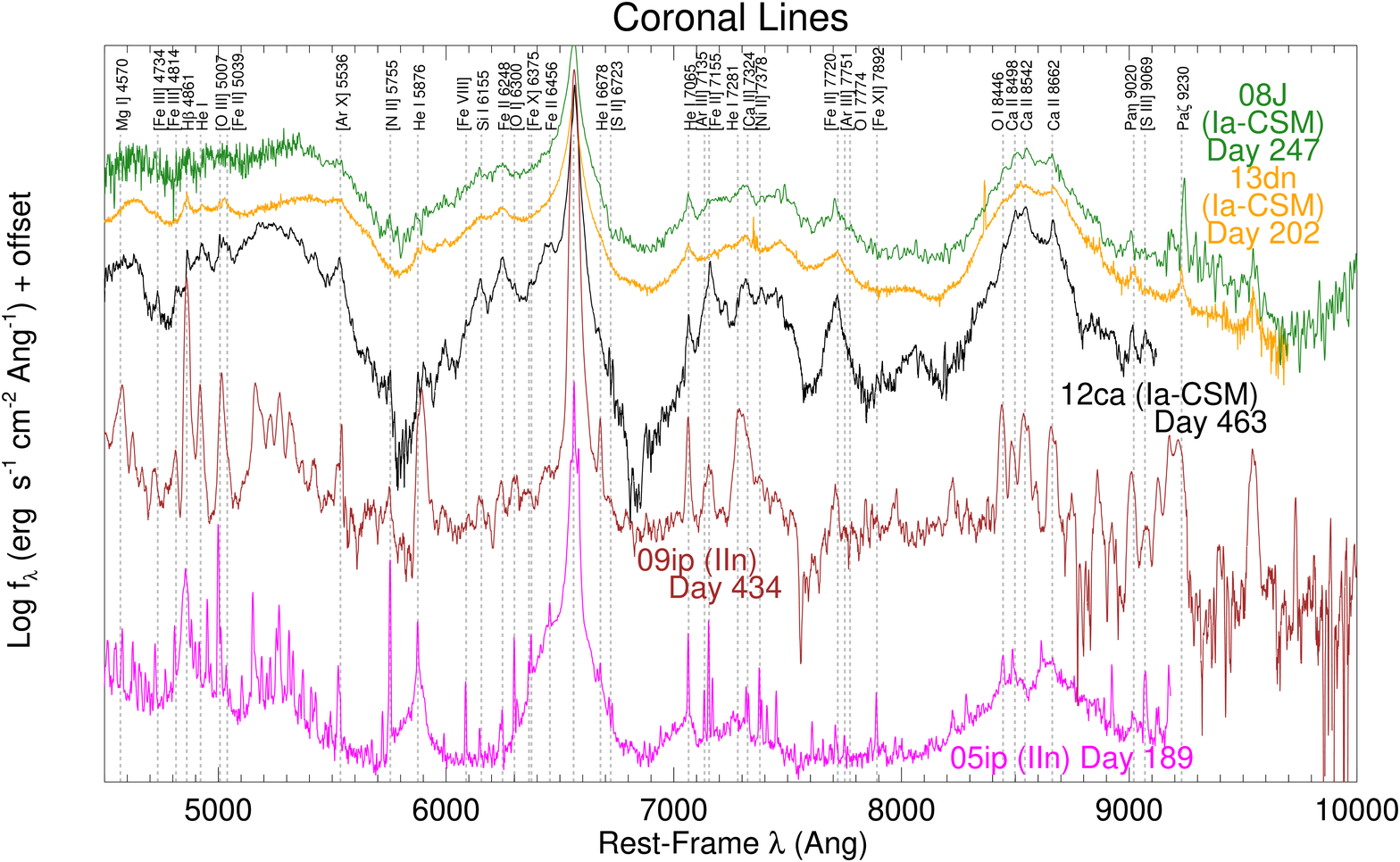}
\caption{Comparison of late-time spectra, including the Type IIn/Ia-CSM SN 2008J (green), the Type Ia-CSM SN 2013dn (orange), the Type Ia-CSM SN 2012ca (black), the Type IIn SN 2009ip (brown), and the Type IIn SN 2005ip (pink).  Some of high-ionisation coronal lines originally identified in SN 2005ip are seen in a number of SNe~Ia-CSM and listed in Table \ref{tab_opt_lines}.
}
\label{fig_foxb} 
\end{figure*}

Section \ref{sec_oxy} suggests alternate line identifications for the blueshifted oxygen.  Compared to SN 2005ip \citep{smith09ip}, Figure \ref{fig_oxy} shows that [\ion{Fe}{II}] emission corresponds to the wavelengths at these blueshifts.  \citet{smith09ip} identify a number of other forbidden iron and relatively high-ionisation lines in SN 2005ip, many of which are more commonly associated with the solar corona (coronal lines; e.g., \citealt{wagner68}), active galactic nuclei (AGNs; e.g., \citealt{filippenko89}), and novae (e.g., \citealt{williams91}), but are rare in SNe.  \citet{smith09ip} point out that no other SN has exhibited such a variety of narrow coronal lines, aside from a few of the stronger ones in SNe~IIn 1988Z and 1995N \citep{turatto93, fransson02}, although a number of these lines are also observed in nearby SN remnants \citep{fesen96}.

Figure \ref{fig_foxb} compares SNe~Ia-CSM 2012ca (black) and 2013dn (orange) to SNe~IIn 2005ip (pink) and 2009ip (brown).  Using line lists made for SN 2005ip \citep{smith09ip} and nearby supernova remnants \citep{fesen96}, we identify a number of high-ionisation lines in SNe 2012ca and 2009ip (also see Table \ref{tab_opt_lines}), including [\ion{S}{III}], [\ion{Ar}{III}], [\ion{Ar}{X}], [\ion{Fe}{VIII}], [\ion{Fe}{X}], and possibly [\ion{Fe}{XI}], although all lines are not quite as prominent in the SNe~Ia-CSM as they are in SN 2005ip. The lines we detect generally correspond to the strongest coronal lines in SN 2005ip, since the underlying continuum likely overwhelms the presences of the weaker lines.

\citet{smith09ip} attribute the narrow ($\sim 120$--240 \kms) coronal lines in SN 2005ip to pre-shock ionisation of the CSM by sustained X-ray emission from ongoing CSI.  The ionisation potentials imply high temperatures up to $\sim2\times10^6$~K.  \citet{smith09ip} go on to explain the lack of any higher-velocity components in these lines with a clumpy or asymmetric CSM, which can arise from various progenitor systems with evolved stars and do not necessarily distinguish between a core-collapse and thermonuclear explosion.  Dense clumps can decelerate the forward shock, giving rise to intermediate-width H$\alpha$~lines, but also leading to efficient cooling and suppression of coronal lines in the post-shock cooling region.  At the same time, the X-rays generated by the CSI can escape along paths without clumps to ionise the pre-shock CSM.

\subsection{Near-Infrared Line Identifications}
\label{sec_ir_spectra}

\begin{figure}
\centering
\includegraphics[width=3.5in]{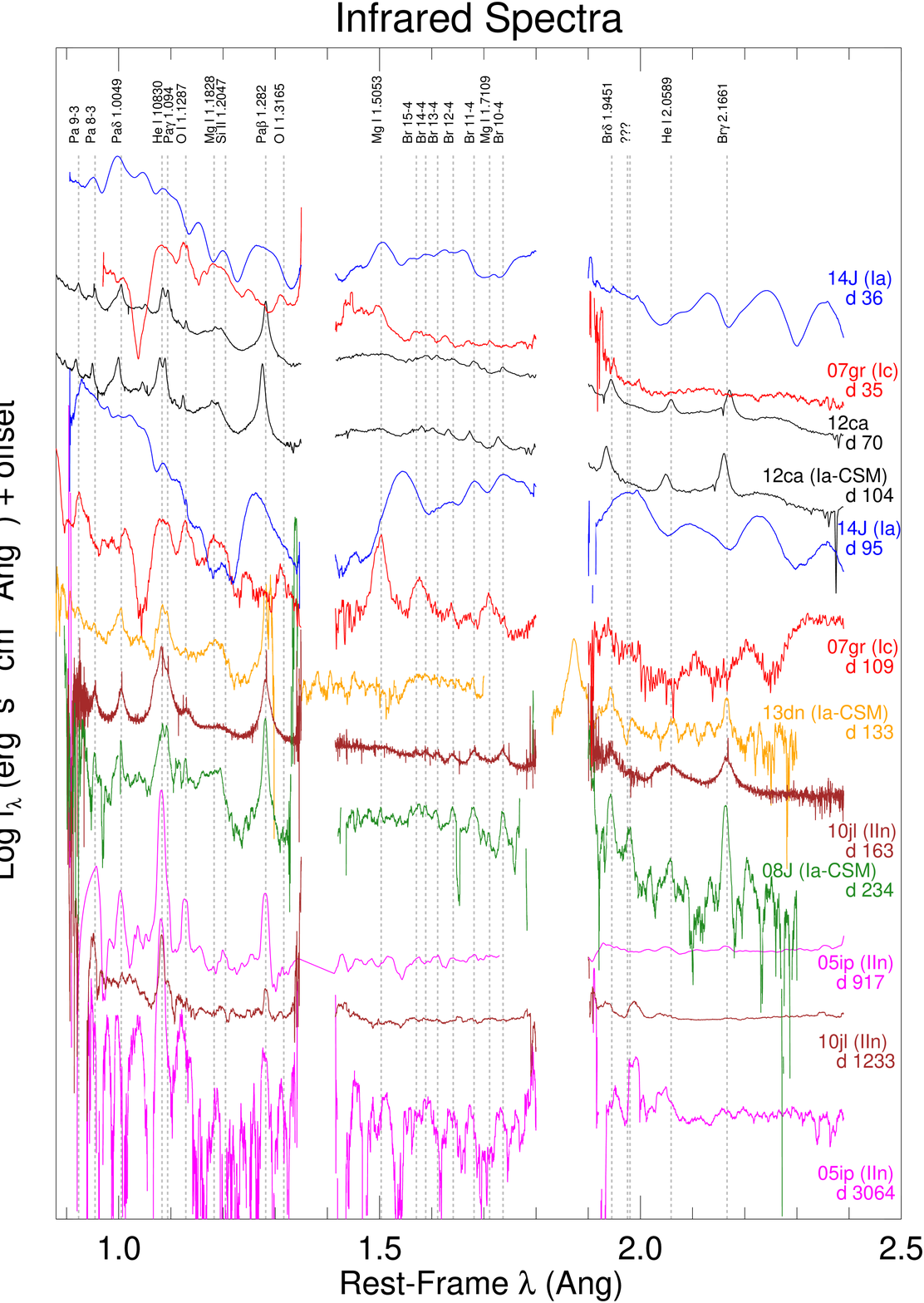}
\caption{Comparison of IR spectra, including the Type IIn/Ia-CSM SN 2008J (green), the Type Ia-CSM SN 2013dn (orange), the Type Ia-CSM SN 2012ca (black), the Type Ic SN 2007gr (red), the Type IIn SN 2010jl (brown), the Type Ia SN 2014J (dark blue), and the Type IIn SN 2005ip (pink).
}
\label{fig_ir_late_Ic} 
\end{figure}

Figure \ref{fig_ir_late_Ic} compares the near-IR spectra of the Type IIn/Ia-CSM SN 2008J (green), the Type Ia-CSM SN 2013dn (orange), the Type Ia-CSM SN 2012ca (black), the Type Ic SN 2007gr (red), the Type IIn SN 2010jl (brown), the Type Ia SN 2014J (dark blue), and the Type IIn SN 2005ip (pink).  We identify spectral features from line lists previously compiled for some well-sampled near-IR spectra, including those of the Type Ic SN 2007gr \citep{hunter09}, the Type IIn SN 1998S \citep{fassia01}, and the \citet{marion09} Type Ia sample (see also Table \ref{tab_ir_lines}).  Not all lines appear in all spectra.  While the underlying continuum from CSI again makes comparisons difficult, we discuss the primary features below.

{\it H:} Both the SNe~Ia-CSM and SNe~IIn near-IR spectra have numerous hydrogen lines.  The more prominant Paschen and Brackett lines are identified throughout the spectrum.  Like other SNe~IIn, the line widths of a few thousand \kms\ are consistent with CSM swept up by the forward shock.  

\begin{table}
\centering
\footnotesize
\caption{Optical Line Identifications \label{tab_opt_lines}}
\begin{tabular}{ l c l c}
\hline
$\lambda$ (\AA) & Line & $\lambda$ (\AA) & Line \\
\hline
4733.93 & [\ion{Fe}{III}] & 7065.19 & \ion{He}{I} \\ 
4813.9 & [\ion{Fe}{III}] & 7135.80 & [\ion{Ar}{III}] \\ 
4861.36 & H$\beta$ & 7155.14 & [\ion{Fe}{II} \\ 
4921.93 & \ion{He}{I} & 7281.35 & \ion{He}{I} \\ 
5039.10 & [\ion{Fe}{II}] & 7323.88 & [\ion{Ca}{II}]\\ 
5536 & [\ion{Ar}{X}] & 7377.83 & [\ion{Ni}{II}] \\ 
5754.59 & [\ion{Ni}{II}] & 7719.9 & [\ion{Fe}{II}] \\ 
5876 blend & \ion{He}{I} & 7751.06 & [\ion{Ar}{III}] \\ 
6087.90 & [\ion{Fe}{VIII}] & 7774 blend & \ion{O}{I} \\ 
6155 & \ion{Si}{I} & 7891.80 & [\ion{Fe}{XI}] \\
6247.56 & \ion{Fe}{II} & 8498.02 & \ion{Ca}{II} \\ 
6374.51 & [\ion{Fe}{X}] & 8542.09 & \ion{Ca}{II} \\ 
6456.38 & \ion{Fe}{II} & 8662.14 & \ion{Ca}{II}\\ 
6560 &  H$\alpha$ & 9020 & Pa$\eta$ \\ 
6678.15 & \ion{He}{I} & 9069.0 & [\ion{S}{III}] \\ 
6723 & [\ion{S}{II}] & 9230 & Pa$\zeta$ \\ 
\hline
\end{tabular}
\end{table}

{\it Helium:} A line at $\sim 1.0830$ \micron\ is detected in SNe 2007gr, 2013dn, 2008J, 2012ca, 2010jl, and 2005ip.  This line is commonly associated with He, but it can be easily confused with \ion{Si}{I} $\lambda$10827.  In fact, \citet{hunter09} attribute this feature in SN 2007gr to a combination of \ion{He}{I}, \ion{Si}{I}, and \ion{Mg}{II}, while the large absorption trough at $\sim 1.04$ \micron\ is attributed to C~I.  Furthermore, at a width of $\sim10^4$~\kms, the line is heavily blended with Pa$\gamma$, and the red wing of the feature bleeds into the blue wing of \ion{O}{I} \citep{borish14}.

A more clearly defined He line is at 2.0589 \micron\ \citep{modjaz09}.  A broad line at this wavelength is also detected in SNe 2013dn, 2008J, 2012ca, 2010jl, and 2005ip, but not in SN 2007gr.  In fact, He is also detected in the optical spectra at 7065 \AA.  Although \citet{silverman13b} note little or no He~I emission in their SN~Ia-CSM sample, their spectra were obtained at earlier epochs when still dominated by an underlying continuum.  Regardless, we see no reason for He to distinguish between a thermonuclear or core-collapse event.  The He, which has similar velocities as the H lines, likely originated in the progenitor system's wind and was accelerated by the forward shock.  A number of scenarios can produce a He wind in the years leading up to the white dwarf explosion \citep[e.g.,][]{lundqvist13}.

{\it \ion{Mg}{I}:} \citet{inserra14} identify \ion{Mg}{I}] $\lambda\lambda$11,300, 15,024, which they use to argue for a core-collapse explosion.  Indeed, these lines (along with \ion{Mg}{I} $\lambda$17,109) are prominent in the Type Ic SN 2007gr.  However, the \ion{Mg}{I} $\lambda\lambda$15,024, 17,109 lines are not detected in any of the SNe~Ia-CSM or SN 2010jl, and the \ion{Mg}{I} $\lambda$11,300 line is relatively weak, if detected at all.  Furthermore, the 1.1300 \micron\ line identification can be easily confused with \ion{O}{I} $\lambda$1.1287 \micron, which we find is more precisely aligned with the features apparent in SNe 2012ca and 2008J.  Compared to Type IIn SNe 2010jl and 2005ip, these lines are more narrow, which likely suggest that they do not originate in the ejecta.  Weakening of the apparent line strength by a strong, underlying continuum must be considered, but the \ion{Mg}{I} lines are some of the strongest lines observed in the Type Ic SN 2007gr and should still be observed in the interacting SNe if present.  Furthermore, Figure \ref{fig_opt_late_lin_300} (bottom) shows a clear detection of \ion{Mg}{I}] $\lambda$4570 in the nebular spectra of SN 1998bw, but not in the SNe~Ia-CSM.  Broad emission features near this wavelength may be confused with Mg, but are more likely the suppressed [\ion{Fe}{III}] discussed in \S \ref{sec_optev}.
 
{\it Oxygen:} Several oxygen lines also reside in the near-IR, including \ion{O}{I} $\lambda\lambda$11,287, 13,165 and \ion{O}{II} $\lambda$21,085.  As expected, these lines are prominently detected in SN 2007gr.  If the 1.1287~\micron\ line is \ion{O}{I} (as we suggest above), it is significantly weaker in SN 2012ca, 2008J, and 2013dn than in SN 2007gr.  This line, however, most likely has contributions from other elements anyway.  A small detection at 1.3165 \micron\ is observed in the day $+$66 spectrum of SN 2012ca, but this might be from the surrounding CSM and is certainly not persistent.  No \ion{O}{I} $\lambda$13,165 or \ion{O}{II} $\lambda$21,085 is detected in SNe 2012ca, 2008J, 2013dn, or even 2010jl.  Furthermore, none of these lines requires a blueshift like that described for the optical spectra of SN 2012ca.  Nonetheless, it should be noted that these near-IR spectra are still dominated by shock-related emission and do not adequately probe nebular lines from the ejecta.  The spectrum of SN 2005ip on day $+$3064, which clearly shows the emergence of broad nebular oxygen lines, underscores this point.

\begin{table}
\centering
\caption{Infrared Line Identifications \label{tab_ir_lines}}
\begin{tabular}{ l c l c}
\hline
$\lambda$ (\micron) & Line & $\lambda$ (\micron) & Line \\
\hline
0.92315 & Pa 9-3 & 1.5053 & \ion{Mg}{I} \\
0.95486 & Pa 8-3 &1.5705 & Br 15-4 \\ 
1.0049 & P$\delta$ & 1.5885 & Br 14-4 \\ 
1.0830 & \ion{He}{I} & 1.6114 & Br 13-4 \\
1.0938 & P$\gamma$ & 1.6412 & Br 12-4 \\ 
1.1287 &  \ion{O}{I} & 1.6811 & Br 11-4 \\
1.1828 & \ion{Mg}{I} & 1.7109 & \ion{Mg}{I}\\
1.2047 & \ion{Mg}{I} & 1.7367 & Br 10-4 \\
1.2818 & P$\beta$ & 1.9451 & Br$\delta$\\ 
1.3165 & \ion{O}{I} & 2.1661 & Br$\gamma$ \\
\hline
\end{tabular}
\end{table}

\section{Discussion}
\label{sec_discussion}

\subsection{Source of the Emission at Late Times}

To interpret the spectra first requires identification of the spectral line origins in the physical SN environment.  Figure \ref{fig_lightcurve} shows that some of the SNe~Ia-CSM and SNe~IIn can be a factor of 100 more luminous than even SN 1991T at late times.  The normal radioactively powered nebular phase of the underlying thermonuclear explosion can therefore only contribute $< 5$\% of the total flux at these epochs.  The majority of the luminosity is instead powered by CSI (and so using the word ``nebular'' is not quite appropriate even at these late epochs).

While the luminosity may be dominated by CSI, the ejecta need not be completely obscured by an opaque shell.  The CSM in many of these systems is likely asymmetric \citep[e.g.,][]{mauerhan14}.  The CDS also continues to expands and can become optically thin over time.  Reprocessing of X-rays and ultraviolet (UV) light generated by the shock interaction can illuminate the ejecta to observable levels.  Furthermore, the ejecta will be excited as they cross the reverse shock at late epochs\citep[e.g.,][]{mauerhan12}.  

Any differences between the spectral lines of normal SNe~Ia and SNe~Ia-CSM/IIn may therefore be caused by any combination of the following reasons.  (1) Stratification of the composition and/or ionisation of the ejecta, since they are illuminated from the outside-in, not from the inside-out.  (2) The source of heat is thermalised UV and X-rays or the crossing of the reverse shock, but not gamma rays from radioactivity, which could also have a strong impact on the ionisation stratification of the ejectal.  (3) Inherent differences in the explosion type and ejecta composition.  Radiative transfer models can provide a more definitive answer, but they are beyond the scope of this paper.

\subsection{[\ion{Fe}{III}] and the Blue ``Quasi-Continuum''}

The broad [\ion{Fe}{III}] and [\ion{Fe}{II}] highlighted in Figure \ref{fig_opt_late_lin_300}, often referred to as a blue ``quasi-continuum,'' is composed of blended iron-group elements (e.g., \citealt{foley07,smith09ip,smith12hw}).  \citet{silverman13b} point to this quasi-continuum in favour of a thermonuclear origin despite weak [\ion{Fe}{III}] when compared to most nebular SNe~Ia (e.g., SN 1991T).  We note, however, that weak [\ion{Fe}{III}] has been observed previously in other SNe~Ia.  Here we refer to the case of SC candidate SN 2009dc, which was believed to have resulted from the explosion of a WD exceeding the Chandrasekhar limit due to high rotation velocities (and also modeled as a ``tamped detonation'' due to merging white dwarfs by \citealt{raskin14}).  SN 2009dc showed suppressed [\ion{Fe}{III}].  

Figure \ref{fig_opt_late_lin_300} highlights the similarities SN 2009dc (pink) shares with the SNe~Ia-CSM (aside from H$\alpha$ of course).  In fact, the [\ion{Fe}{III}] emission is still stronger in SN 2009dc than in the SNe~Ia-CSM (particularly on day $+$164), but significantly less than in the overluminous SN 1991T.  \citet{taubenberger13} attribute the smaller [\ion{Fe}{III}]/[\ion{Fe}{II}] ratio in SN 2009dc to a low-ionisation state during the nebular phase owing to high central ejecta densities (and therefore, enhanced recombination).  In the case of SN 2009dc, these high densities may be a result of the low ejecta expansion velocities observed in SC SNe \citep[e.g.,][]{silverman11, taubenberger11}, whereas for the SNe~Ia-CSM the high densities likely result from decelerated ejecta by the dense CSM.  \citet{taubenberger13} go on to show that the low-ionisation state would also be consistent with the detection of [\ion{Ca}{II}] $\lambda\lambda$7291, 7324, since the first and second ionisation potentials of Ca are lower than those of iron.  Indeed, Figure \ref{fig_opt_late_lin_300} shows that [\ion{Ca}{II}] is detected in both the SC and SNe~Ia-CSM.

Admittedly, such a blue ``quasi-continuum'' has also been observed in other interacting SNe, including the SNe~Ibn 2006jc \citep{foley07} and 2011hw \citep{smith12hw} and the SN~IIn 2005ip \citep{smith09ip}.  As described for the cases of SNe~Ibn 2006jc and 2011hw \citep{smith12hw}, which have lost a majority of their H envelopes, a low H abundance will place the burden of the radiative cooling from the CSI on the Fe emission lines.  In other words, the Fe lines may very well originate in the ejecta, but the strength of the lines {\it may} be an indicator of the H, rather than the Fe, abundance.

Figure \ref{fig_opt_late_lin_300} compares the SNe Ia-CSM 2013dn (orange) and 2012ca (black) to the SN Ibn 2006jc (green), the SN~IIn 2009ip (brown), and the SC SN 2009dc (pink).  While there are some similarities in the shapes of the ``quasi-continuum,'' there are several important differences to the overall spectra.  First, the hydrogen features in SNe 2013dn and 2012ca are significantly stronger than those in SN 2006jc, suggesting that the H abundance is not particularly low in the SNe Ia-CSM.  The H line strength may not directly indicate the total gas mass, but the light curves in Figure \ref{fig_lightcurve} show that the luminosity from CSM interaction is significantly stronger and more extended in SNe 2012ca and 2013dn than in either SN 2006jc or SN 2005ip.  The total mass of gas, composed primarily of H, is therefore larger in these SNe Ia-CSM.  Second, the [\ion{Fe}{III}] lines are present in SNe 2013dn and 2009dc, and even identifiable in SN 2012ca, but seemingly absent in SN 2006jc.  Third, the Mg~I] $\lambda$4570 feature in SN 1998bw is detected in SN 2006jc, but not in SNe 2009dc, 2013dn, or 2012ca.  We are surely limited by the epochs of data we have to compare SN 2009dc to SN 2012ca since the [\ion{Fe}{III}] becomes only more suppressed over time in SN 2009dc.  



\subsection{The Nature of the Type Ia-CSM Progenitor}

\citet{inserra14} suggest the SNe~Ia-CSM match well with SN 1998bw, but we find this comparison difficult to reconcile with our data.  Namely, [\ion{O}{I}] $\lambda$6300 in SN 1998bw is too strong (as is [\ion{Ca}{II}]), particularly considering we do not detect any significant oxygen in our SN~Ia-CSM spectra (see \S \ref{sec_oxy}).  We also do not detect any carbon or magnesium.  The SNe~Ia CSM feature at $\sim 4600$--4700~\AA\ is more likely suppressed [\ion{Fe}{III}] than the \ion{Mg}{I}] found in SN 1998bw.  The weak [\ion{Fe}{III}] and strong [\ion{Ca}{II}] in the nebular spectra of the SNe~Ia-CSM are most similar to the SC SN 2009dc, which was modeled by a low-ionisation state owing to high densities that we would also expect in the SNe~Ia-CSM.

Furthermore, Figure \ref{fig_lightcurve} shows the bolometric light curve of several SNe~Ia-CSM.  All SN~Ia-CSM peak magnitudes are $> -19$, and the total integrated radiated luminosity output is a few $\times10^{50}$~erg, which is still consistent with a thermonuclear explosion of a white dwarf ($\sim10^{51}$~erg) but requires a high conversion efficiency of kinetic energy into radiation ($\epsilon \approx 0.5$).  While such conversion efficiencies are quite high, they are certainly possible \citep[e.g.,][]{marle10}.  The Type Ia IR signatures are less apparent.  

\begin{figure}
\centering
\includegraphics[width=3.5in]{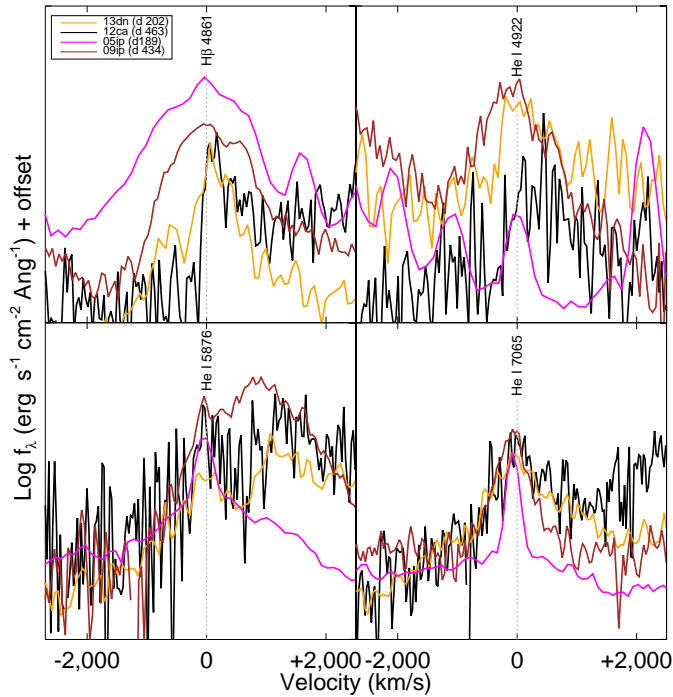}
\vspace{0.1in}
\caption{Velocity profiles of several H/He lines, including the Type Ia-CSM SN 2013dn (orange), the Type Ia-CSM SN 2012ca (black), the Type IIn SN 2009ip (brown), and the Type IIn SN 2005ip (pink).  Vertical dashed lines identify prominent spectral lines, although it should be noted that not all lines appear in all spectra.
}
\label{fig_he_vel} 
\end{figure}

With a thermonuclear progenitor in mind, we reconsider Figure \ref{fig_foxb}, which highlights strong iron lines (e.g., $\lambda\lambda$6248, 7155, 7720) in the spectra of SNe 2012ca, 2009ip, 2013dn, and 2008J.  Figures \ref{fig_oxy} plots the velocity profiles for these iron lines.  Compared to SN 2005ip, both the SNe~Ia-CSM and the SN~IIn 2009ip have broader line profiles ($\ge 1000$ \kms).  If associated with the pre-shock wind, as suggested in the case of SN 2005ip, the higher velocities could potentially imply progenitor scenarios that differ from the massive-star progenitors with low speed winds (i.e., LBVs) proposed for most SNe~IIn.  These velocities, however, are not consistent with the more narrow hydrogen and helium ($\le 1000$ \kms; see Figure \ref{fig_he_vel}), which are associated with the pre-shock CSM.  Instead, these broad iron lines more likely originate in the post-shock cooling region or in the ejecta, which would be more consistent with SNe~Ia.

The broad iron lines in SN 2009ip may be surprising given it has a progenitor that has most certainly been identified as a massive star (i.e., core collapse; see \S \ref{sec_obs}).  Of course, X-rays produced by the CSI may excite even solar abundances of iron in post-shock CSM.  In this case, these iron features would not be a useful discriminator.  Furthermore, Figure \ref{fig_opt_late_lin_300} (bottom) compares a day $+$260 spectrum of the interacting Type IIn SN 2009ip (brown) to the SNe~Ia-CSM.  While SN 2009ip does exhibit some evidence for iron emission shortward of 5500~\AA, it is significantly weaker than the other thermonuclear events.  This fact alone is an important differentiator between the SNe~IIn and SNe~Ia-CSM in our sample.

Despite our discussion about the oxygen misidentifications in \S \ref{sec_oxy}, we note that we do detect some oxygen $\lambda\lambda$5007, 6300, 8446 in the nebular spectra of both SNe~Ia-CSM 2013dn, 2012ca and SNe~IIn 2009ip, 2005ip.  The oxygen emission, however, is relatively weak and narrow, and the broader lines at 8446 \AA\ are more likely a result of Ly$\alpha$ pumping (which scales linearly with density) and not recombination.  In this case, the lack of intermediate-mass elements in SN 2009ip and 2005ip may be more of the surprise.  These results have been used to argue against a terminal core collapse in SN 2009ip already \citep{fraser13b}, but \citet{smith14ip} point out that even their late-time spectra were dominated by CSI.  Even if the ejecta are illuminated, the illumination comes from the outside shocks and the intermediate-mass elements may be hidden deep in the ejecta or the lines may be weak.  Indeed, the lack of nucleosynthetic signatures in SNe~IIn is a common trait (e.g., SNe 2005ip and 1998S) that results from the formation of an optically thick shell from CSI that obscures emission from the ejecta \citep{mauerhan12}.  Late-time spectra of SNe~IIn only begin to show evidence for broad oxygen emission if the ejecta are still bright enough once the CSI sufficiently fades and/or the ejecta begin to cross the reverse shock.  Therefore, this comparison may be less relevant than previously suggested.

\section{Conclusions and Future Work}
\label{sec_conclusion}

We find the SN~Ia-CSM subclass to be more consistent with a thermonuclear explosion than a core-collapse event.  Specifically, the spectra do not showing evidence for intermediate-mass elements and do exhibit broad iron features, a low [\ion{Fe}{III}]/[\ion{Fe}{II}] ratio most similar to the super-Chandrasekhar-mass candidate SN 2009dc, and a total bolometric energy output that does not exceed $10^{51}$~erg.  Nonetheless, there is still some ambiguity as to the origin of the blue ``quasi-continuum.''  Larger samples comparing these features as a function of bolometric luminosity and hydrogen line strength will be necessary to further disentangle the origin of the quasi-continuum.  This work also highlights the need for robust radiative transfer models for comparison.

While the case for a thermonuclear origin appears to have gained a more robust foothold, the companion's properties still remain relatively unknown.  Furthermore, the ability for the companion to undergo such high mass loss (i.e., $>10^{-3}$~\msolar; see \citealt{silverman13b}, and references therein) remains poorly understood, although the presence of a binary likely has a role \citep[e.g.,][]{smith14lbv}.  Future multi-wavelength observations will be needed to probe the CSM characteristics, trace the CSI, constrain the progenitor mass-loss history, and identify late-time heating mechanisms of warm dust.
\newline

Insightful discussions were shared with many at the Aspen Center for Physics, including Ryan Foley, Ryan Chornock, and Craig Wheeler.  This work was supported in part by NSF Grant No. PHYS-1066293 and the hospitality of the Aspen Center for Physics.  Some of the data presented herein were obtained at the W. M. Keck Observatory, which is operated as a scientific partnership among the California Institute of Technology, the University of California, and NASA; the observatory was made possible by the generous financial support of the W. M. Keck Foundation. We are grateful to the staffs of the Lick and Keck Observatory for their assistance with the observations, and thank the RATIR instrument team and the staff of the Observatorio Astrono\'mico Nacional on Sierra San Pedro Ma\'rtir. RATIR is a collaboration between the University of California, the Universidad Nacional Autono\'ma de Me\'xico, NASA Goddard Space Flight Center, and Arizona State University, benefiting from the loan of an H2RG detector from Teledyne Scientific and Imaging. RATIR, the automation of the Harold L. Johnson Telescope of the Observatorio Astrono\'mico Nacional on Sierra San Pedro Ma\'rtir, and the operation of both are funded by the partner institutions and through NASA grants NNX09AH71G, NNX09AT02G, NNX10AI27G, and NNX12AE66G, CONACyT grant INFR-2009-01-122785, UNAM PAPIIT grant IN113810, and a UC MEXUS-CONACyT grant.  J.M.S. is supported by an NSF Astronomy and Astrophysics Postdoctoral Fellowship under award AST--1302771. A.V.F.'s supernova group at UC Berkeley received support through NSF grant AST--1211916, the TABASGO Foundation, Gary and Cynthia Bengier, the Richard and Rhoda Goldman Fund, and the Christopher R. Redlich Fund.


\bibliographystyle{apj2}
\bibliography{references}

\begin{thebibliography}{}
\expandafter\ifx\csname natexlab\endcsname\relax\def\natexlab#1{#1}\fi

\bibitem[{Ahn {et~al.}(2012)Ahn, Alexandroff, Prieto, Anderson, Anderton,
  Andrews, Aubourg, Bailey, Balbinot, Barnes, Bautista, Beers, Beifiori,
  Berlind, Bhardwaj, Bizyaev, Blake, Blanton, Blomqvist, Bochanski, Bolton,
  Borde, Bovy, Brandt, Brinkmann, Brown, Brownstein, Bundy, Busca, Carithers,
  Carnero, Carr, Casetti-Dinescu, Chen, Chiappini, Comparat, Connolly, Crepp,
  Cristiani, Croft, Cuesta, da~Costa, Davenport, Dawson, de~Putter, Lee,
  Delubac, Dhital, Ealet, Ebelke, Edmondson, Eisenstein, Escoffier, Esposito,
  Evans, Fan, Castell{\'a}, Alvar, Ferreira, Ak, Finley, Fleming, Font-Ribera,
  Frinchaboy, Garc{\'\i}a-Hern{\'a}ndez, P{\'e}rez, Ge, G{\'e}nova-Santos,
  Gillespie, Girardi, Hern{\'a}ndez, Grebel, Gunn, Guo, Haggard, Hamilton,
  Harris, Hawley, Hearty, Ho, Hogg, Holtzman, Honscheid, Huehnerhoff, Ivans,
  Ivezi{\'c}, Jacobson, Jiang, Johansson, Johnson, Kauffmann, Kirkby,
  Kirkpatrick, Klaene, Knapp, Kneib, Goff, Leauthaud, Lee, Lee, Long, Loomis,
  Lucatello, Lundgren, Lupton, Ma, Ma, MacDonald, Mack, Mahadevan, Maia,
  Majewski, Makler, Malanushenko, Malanushenko, Manchado, Mandelbaum, Manera,
  Maraston, Margala, Martell, McBride, McGreer, McMahon, M{\'e}nard, Meszaros,
  Miralda-Escud{\'e}, Montero-Dorta, Montesano, Morrison, Muna, Munn, Murayama,
  Myers, Neto, Nguyen, Nichol, Nidever, Noterdaeme, Nuza, Ogando, Olmstead,
  Oravetz, Owen, Padmanabhan, Palanque-Delabrouille, Pan, Parejko, Parihar,
  P{\^a}ris, Pattarakijwanich, Pepper, Percival, P{\'e}rez-Fournon,
  P{\'e}rez-R{\`a}fols, Petitjean, Pforr, Pieri, Pinsonneault, de~Mello, Prada,
  Price-Whelan, Raddick, Rebolo, Rich, Richards, Robin, Rocha-Pinto, Rockosi,
  Roe, Ross, Ross, Rossi, Rubi{\~n}o-Martin, Samushia, Almeida, S{\'a}nchez,
  Santiago, Sayres, Schlegel, Schlesinger, Schmidt, Schneider, Schultheis,
  Schwope, Sc{\'o}ccola, Seljak, Sheldon, Shen, Shu, Simmerer, Simmons, Skibba,
  Skrutskie, Slosar, Sobreira, Sobeck, Stassun, Steele, Steinmetz, Strauss,
  Streblyanska, Suzuki, Swanson, Tal, Thakar, Thomas, Thompson, Tinker,
  Tojeiro, Tremonti, Maga{\~n}a, Verde, Viel, Vikas, Vogt, Wake, Wang, Weaver,
  Weinberg, Weiner, West, White, Wilson, Wisniewski, Wood-Vasey, Yanny,
  Y{\`e}che, York, Zamora, Zasowski, Zehavi, Zhao, Zheng, Zhu, \& Zinn}]{ahn12}
Ahn, C.~P., Alexandroff, R., Prieto, C.~A., {et~al.} 2012, ApJS, 203, 21

\bibitem[{Aldering {et~al.}(2006)Aldering, Antilogus, Bailey, Baltay, Bauer,
  Blanc, Bongard, Copin, Gangler, Gilles, Kessler, Kocevski, Lee, Loken,
  Nugent, Pain, P{\'e}contal, Pereira, Perlmutter, Rabinowitz, Rigaudier,
  Scalzo, Smadja, Thomas, Wang, Weaver, \& Factory}]{aldering06}
Aldering, G., Antilogus, P., Bailey, S., {et~al.} 2006, ApJ, 650, 510

\bibitem[{Benetti {et~al.}(2006)Benetti, Cappellaro, Turatto, Taubenberger,
  Harutyunyan, \& Valenti}]{benetti06}
Benetti, S., Cappellaro, E., Turatto, M., {et~al.} 2006, ApJL, 653, L129

\bibitem[{Boles {et~al.}(2005)Boles, Nakano, \& Itagaki}]{boles05}
Boles, T., Nakano, S., \& Itagaki, K. 2005, CBET, 275, 1

\bibitem[{Borish {et~al.}(2014)Borish, Huang, Chevalier, Breslauer, Kingery, \&
  Privon}]{borish14}
Borish, H.~J., Huang, C., Chevalier, R.~A., {et~al.} 2014, arXiv: 1406.5531

\bibitem[{Bowers {et~al.}(1997)Bowers, Meikle, Geballe, Walton, Pinto, Dhillon,
  Howell, \& Harrop-Allin}]{bowers97}
Bowers, E. J.~C., Meikle, W. P.~S., Geballe, T.~R., {et~al.} 1997, MNRAS, 290,
  663

\bibitem[{Branch {et~al.}(2008)Branch, Jeffery, Parrent, Baron, Troxel,
  Stanishev, Keithley, Harrison, \& Bruner}]{branch08}
Branch, D., Jeffery, D.~J., Parrent, J., {et~al.} 2008, PASP, 120, 135

\bibitem[{Butler {et~al.}(2012)Butler, Klein, Fox, Lotkin, Bloom, Prochaska,
  Ramirez-Ruiz, de~Diego, Georgiev, Gonz{\'a}lez, Lee, Richer, Rom{\'a}n,
  Watson, Gehrels, Kutyrev, Bernstein, Alvarez, Cese{\~n}a, Clark, Colorado,
  C{\'o}rdova, Farah, Garc{\'\i}a, Guisa, Herrera, Lazo, L{\'o}pez, Luna,
  Mart{\'\i}nez, Murillo, Murillo, N{\'u}{\~n}ez, Pedrayes, Quir{\'o}s, Ochoa,
  Sierra, Moseley, Rapchun, Robinson, Samuel, \& Sparr}]{butler12}
Butler, N., Klein, C., Fox, O., {et~al.} 2012, Proc. of the SPIE, 8446, 10

\bibitem[{Cushing {et~al.}(2004)Cushing, Vacca, \& Rayner}]{cushing04}
Cushing, M.~C., Vacca, W.~D., \& Rayner, J.~T. 2004, PASP, 116, 362

\bibitem[{Deng {et~al.}(2004)Deng, Kawabata, Ohyama, Nomoto, Mazzali, Wang,
  Jeffery, Iye, Tomita, \& Yoshii}]{deng04}
Deng, J., Kawabata, K.~S., Ohyama, Y., {et~al.} 2004, ApJ, 605, L37

\bibitem[{Dilday {et~al.}(2012)Dilday, Howell, Cenko, Silverman, Nugent,
  Sullivan, Ben-ami, Bildsten, Bolte, Endl, Filippenko, Gnat, Horesh, Hsiao,
  Kasliwal, Kirkman, Maguire, Marcy, Moore, Pan, Parrent, Podsiadlowski,
  Quimby, Sternberg, Suzuki, Tytler, Xu, Bloom, Gal-Yam, Hook, Kulkarni, Law,
  Ofek, Polishook, \& Poznanski}]{dilday12}
Dilday, B., Howell, D.~A., Cenko, S.~B., {et~al.} 2012, Science, 337, 942

\bibitem[{Drake {et~al.}(2013)Drake, Djorgovski, Graham, Mahabal, Williams,
  Prieto, Catelan, Larson, Christensen, Brimacombe, Benetti, Pastorello,
  Cappellaro, Tomasella, Ochner, Turatto, \& Harutyunyan}]{drake13}
Drake, A.~J., Djorgovski, S.~G., Graham, M.~J., {et~al.} 2013, CBET, 3570, 1

\bibitem[{Drescher {et~al.}(2012)Drescher, Parker, \& Brimacombe}]{drescher12}
Drescher, C., Parker, S., \& Brimacombe, J. 2012, CBET, 3101, 1

\bibitem[{Faber {et~al.}(2003)Faber, Phillips, Kibrick, Alcott, Allen, Burrous,
  Cantrall, Clarke, Coil, Cowley, Davis, Deich, Dietsch, Gilmore, Harper,
  Hilyard, Lewis, McVeigh, Newman, Osborne, Schiavon, Stover, Tucker, Wallace,
  Wei, Wirth, \& Wright}]{faber03}
Faber, S.~M., Phillips, A.~C., Kibrick, R.~I., {et~al.} 2003, Proc. of SPIE,
  4841, 1657

\bibitem[{Fassia {et~al.}(2001)Fassia, Meikle, Chugai, Geballe, Lundqvist,
  Walton, Pollacco, Veilleux, Wright, Pettini, Kerr, Puchnarewicz, Puxley,
  Irwin, Packham, Smartt, \& Harmer}]{fassia01}
Fassia, A., Meikle, W. P.~S., Chugai, N., {et~al.} 2001, MNRAS, 325, 907

\bibitem[{Fesen \& Hurford(1996)}]{fesen96}
Fesen, R.~A., \& Hurford, A.~P. 1996, ApJS, 106, 563

\bibitem[{Filippenko(1982)}]{filippenko82}
Filippenko, A.~V. 1982, PASP, 94, 715

\bibitem[{Filippenko(1989)}]{filippenko89}
---. 1989, AJ, 97, 726

\bibitem[{Filippenko(1997)}]{filippenko97}
---. 1997, ARA\&A, 35, 309

\bibitem[{Foley {et~al.}(2011)Foley, Berger, Fox, Levesque, Challis, Ivans,
  Rhoads, \& Soderberg}]{foley11}
Foley, R.~J., Berger, E., Fox, O., {et~al.} 2011, ApJ, 732, 32

\bibitem[{Foley {et~al.}(2007)Foley, Smith, Ganeshalingam, Li, Chornock, \&
  Filippenko}]{foley07}
Foley, R.~J., Smith, N., Ganeshalingam, M., {et~al.} 2007, ApJ, 657, L105

\bibitem[{Foley {et~al.}(2003)Foley, Papenkova, Swift, Filippenko, Li, Mazzali,
  Chornock, Leonard, \& van Dyk}]{foley03}
Foley, R.~J., Papenkova, M.~S., Swift, B.~J., {et~al.} 2003, PASP, 115, 1220

\bibitem[{Fox {et~al.}(2009)Fox, Skrutskie, Chevalier, Kanneganti, Park,
  Wilson, Nelson, Amirhadji, Crump, Hoeft, Provence, Sargeant, Sop, Tea,
  Thomas, \& Woolard}]{fox09}
Fox, O.~D., Skrutskie, M.~F., Chevalier, R.~A., {et~al.} 2009, ApJ, 691, 650

\bibitem[{Fox {et~al.}(2012)Fox, Kutyrev, Rapchun, Klein, Butler, Bloom,
  de~Diego, Farah, Gehrels, Georgiev, Gonz{\'a}lez, Lee, Loose, Lotkin,
  Moseley, Prochaska, Ramirez-Ruiz, Richer, Robinson, Rom{\'a}n-Z{\'u}{\~n}iga,
  Samuel, Sparr, \& Watson}]{fox12}
Fox, O.~D., Kutyrev, A.~S., Rapchun, D.~A., {et~al.} 2012, Proc. of SPIE, 8453,
  59

\bibitem[{Fransson {et~al.}(2002)Fransson, Chevalier, Filippenko, Leibundgut,
  Barth, Fesen, Kirshner, Leonard, Li, Lundqvist, Sollerman, \& van
  Dyk}]{fransson02}
Fransson, C., Chevalier, R.~A., Filippenko, A.~V., {et~al.} 2002, ApJ, 572, 350

\bibitem[{Fraser {et~al.}(2013{\natexlab{a}})Fraser, Kotak, Pastorello,
  Benetti, Inserra, Smartt, Taubenberger, Hachinger, Elias-Rosa, Garoffolo,
  Walker, Valenti, Smith, Young, Sullivan, Gal-Yam, \& Yaron}]{fraser13a}
Fraser, M., Kotak, R., Pastorello, A., {et~al.} 2013{\natexlab{a}}, ATEL, 4953,
  1

\bibitem[{Fraser {et~al.}(2013{\natexlab{b}})Fraser, Inserra, Jerkstrand,
  Kotak, Pignata, Benetti, Botticella, Bufano, Childress, Mattila, Pastorello,
  Smartt, Turatto, Yuan, Anderson, Bayliss, Bauer, Chen, Bur{\'o}n, Gal-Yam,
  Haislip, Knapic, Guillou, Marchi, Mazzali, Molinaro, Moore, Reichart,
  Smareglia, Smith, Sternberg, Sullivan, Tak{\'a}ts, Tucker, Valenti, Yaron,
  Young, \& Zhou}]{fraser13b}
Fraser, M., Inserra, C., Jerkstrand, A., {et~al.} 2013{\natexlab{b}}, MNRAS,
  433, 1312

\bibitem[{Germany {et~al.}(2000)Germany, Reiss, Sadler, Schmidt, \&
  Stubbs}]{germany00}
Germany, L.~M., Reiss, D.~J., Sadler, E.~M., Schmidt, B.~P., \& Stubbs, C.~W.
  2000, ApJ, 533, 320

\bibitem[{Graham {et~al.}(2014)Graham, Sand, Valenti, Howell, Parrent, Halford,
  Zaritsky, Bianco, Rest, \& Dilday}]{graham14}
Graham, M.~L., Sand, D.~J., Valenti, S., {et~al.} 2014, ApJ, 787, 163

\bibitem[{Hamuy {et~al.}(2003)Hamuy, Phillips, Suntzeff, Maza, Gonz{\'a}lez,
  Roth, Krisciunas, Morrell, Green, Persson, \& McCarthy}]{hamuy03b}
Hamuy, M., Phillips, M.~M., Suntzeff, N.~B., {et~al.} 2003, Nature, 424, 651

\bibitem[{Herter {et~al.}(2008)Herter, Henderson, Wilson, Matthews, Rahmer,
  Bonati, Muirhead, Adams, Lloyd, Skrutskie, Moon, Parshley, Nelson,
  Martinache, \& Gull}]{herter08}
Herter, T.~L., Henderson, C.~P., Wilson, J.~C., {et~al.} 2008, Proc. of SPIE,
  7014, 30

\bibitem[{Horne(1986)}]{horne86}
Horne, K. 1986, PASP, 98, 609

\bibitem[{Hunter {et~al.}(2009)Hunter, Valenti, Kotak, Meikle, Taubenberger,
  Pastorello, Benetti, Stanishev, Smartt, Trundle, Arkharov, Bufano,
  Cappellaro, Carlo, Dolci, Elias-Rosa, Frandsen, Fynbo, Hopp, Larionov,
  Laursen, Mazzali, Navasardyan, Ries, Riffeser, Rizzi, Tsvetkov, Turatto, \&
  Wilke}]{hunter09}
Hunter, D.~J., Valenti, S., Kotak, R., {et~al.} 2009, A{\&}A, 508, 371

\bibitem[{Inserra {et~al.}(2012)Inserra, Smartt, Valenti, Pastorello,
  Cappellaro, Benetti, Benitez-Herrera, Taubenberger, Sullivan, \&
  Scalzo}]{inserra12}
Inserra, C., Smartt, S.~J., Valenti, S., {et~al.} 2012, CBET, 3101, 2

\bibitem[{Inserra {et~al.}(2014)Inserra, Smartt, Scalzo, Fraser, Pastorello,
  Childress, Pignata, Jerkstrand, Kotak, Benetti, Valle, Gal-Yam, Mazzali,
  Smith, Sullivan, Valenti, Yaron, Young, \& Reichart}]{inserra14}
Inserra, C., Smartt, S.~J., Scalzo, R., {et~al.} 2014, MNRAS Letters, 437, L51

\bibitem[{Kelson(2003)}]{kelson03}
Kelson, D.~D. 2003, PASP, 115, 688

\bibitem[{Leloudas {et~al.}(2013)Leloudas, Hsiao, Johansson, Maeda, Moriya,
  Nordin, Petrushevska, Silverman, Sollerman, Stritzinger, Taddia, \&
  Xu}]{leloudas13}
Leloudas, G., Hsiao, E.~Y., Johansson, J., {et~al.} 2013, arXiv:1306.1549,
  1306, 1549

\bibitem[{Levesque {et~al.}(2014)Levesque, Stringfellow, Ginsburg, Bally, \&
  Keeney}]{levesque14}
Levesque, E.~M., Stringfellow, G.~S., Ginsburg, A.~G., Bally, J., \& Keeney,
  B.~A. 2014, AJ, 147, 23

\bibitem[{Lundqvist {et~al.}(2013)Lundqvist, Mattila, Sollerman, Kozma, Baron,
  Cox, Fransson, Leibundgut, \& Spyromilio}]{lundqvist13}
Lundqvist, P., Mattila, S., Sollerman, J., {et~al.} 2013, MNRAS, 435, 329

\bibitem[{Maoz {et~al.}(2013)Maoz, Mannucci, \& Nelemans}]{maoz13}
Maoz, D., Mannucci, F., \& Nelemans, G. 2013, arXiv:1312.0628

\bibitem[{Margutti {et~al.}(2014)Margutti, Milisavljevic, Soderberg, Chornock,
  Zauderer, Murase, Guidorzi, Sanders, Kuin, Fransson, Levesque, Chandra,
  Berger, Bianco, Brown, Challis, Chatzopoulos, Cheung, Choi, Chomiuk, Chugai,
  Contreras, Drout, Fesen, Foley, Fong, Friedman, Gall, Gehrels, Hjorth, Hsiao,
  Kirshner, Im, Leloudas, Lunnan, Marion, Martin, Morrell, Neugent, Omodei,
  Phillips, Rest, Silverman, Strader, Stritzinger, Szalai, Utterback, Vinko,
  Wheeler, Arnett, Campana, Chevalier, Ginsburg, Kamble, Roming, Pritchard, \&
  Stringfellow}]{margutti14}
Margutti, R., Milisavljevic, D., Soderberg, A.~M., {et~al.} 2014, ApJ, 780, 21

\bibitem[{Marion {et~al.}(2009)Marion, H{\"o}flich, Gerardy, Vacca, Wheeler, \&
  Robinson}]{marion09}
Marion, G.~H., H{\"o}flich, P., Gerardy, C.~L., {et~al.} 2009, AJ, 138, 727

\bibitem[{Matheson {et~al.}(2000)Matheson, Filippenko, Ho, Barth, \&
  Leonard}]{matheson00}
Matheson, T., Filippenko, A.~V., Ho, L.~C., Barth, A.~J., \& Leonard, D.~C.
  2000, AJ, 120, 1499

\bibitem[{Mauerhan \& Smith(2012)}]{mauerhan12}
Mauerhan, J., \& Smith, N. 2012, MNRAS, 424, 2659

\bibitem[{Mauerhan {et~al.}(2014)Mauerhan, Williams, Smith, Smith, Filippenko,
  Hoffman, Milne, Leonard, Clubb, Fox, \& Kelly}]{mauerhan14}
Mauerhan, J., Williams, G.~G., Smith, N., {et~al.} 2014, MNRAS, 442, 1166

\bibitem[{Mauerhan {et~al.}(2013)Mauerhan, Smith, Filippenko, Blanchard,
  Blanchard, Casper, Cenko, Clubb, Cohen, Fuller, Li, \&
  Silverman}]{mauerhan13}
Mauerhan, J.~C., Smith, N., Filippenko, A.~V., {et~al.} 2013, MNRAS, 749

\bibitem[{Maza {et~al.}(2009)Maza, Hamuy, Antezana, Gonzalez, Lopez, Silva,
  Folatelli, Iturra, Cartier, Forster, Marchi, Rojas, Pignata, Conuel,
  Reichart, Ivarsen, Haislip, Crain, Foster, Nysewander, \& LaCluyze}]{maza09}
Maza, J., Hamuy, M., Antezana, R., {et~al.} 2009, CBET, 1928, 1

\bibitem[{Milisavljevic {et~al.}(2012)Milisavljevic, Fesen, Chevalier,
  Kirshner, Challis, \& Turatto}]{milisavljevic12}
Milisavljevic, D., Fesen, R., Chevalier, R., {et~al.} 2012, ApJ, 751, 25

\bibitem[{Miller \& Stone(1993)}]{miller93}
Miller, J.~S., \& Stone, R. P.~S. 1993, Lick Obs. Tech. Rep. 66

\bibitem[{Modjaz {et~al.}(2005)Modjaz, Kirshner, Challis, \&
  Calkins}]{modjaz05}
Modjaz, M., Kirshner, R., Challis, P., \& Calkins, M. 2005, IAU Circ., 8628, 2

\bibitem[{Modjaz {et~al.}(2009)Modjaz, Li, Butler, Chornock, Perley, Blondin,
  Bloom, Filippenko, Kirshner, Kocevski, Poznanski, Hicken, Foley,
  Stringfellow, Berlind, y~Navascues, Blake, Bouy, Brown, Challis, Chen,
  de~Vries, Dufour, Falco, Friedman, Ganeshalingam, Garnavich, Holden,
  Illingworth, Lee, Liebert, Marion, Olivier, Prochaska, Silverman, Smith,
  Starr, Steele, Stockton, Williams, \& Wood-Vasey}]{modjaz09}
Modjaz, M., Li, W., Butler, N., {et~al.} 2009, ApJ, 702, 226

\bibitem[{Oke {et~al.}(1995)Oke, Cohen, Carr, Cromer, Dingizian, Harris,
  Labrecque, Lucinio, Schaal, Epps, \& Miller}]{oke95}
Oke, J.~B., Cohen, J.~G., Carr, M., {et~al.} 1995, PASP, 107, 375

\bibitem[{Parrent {et~al.}(2014)Parrent, Friesen, \& Parthasarathy}]{parrent14}
Parrent, J., Friesen, B., \& Parthasarathy, M. 2014, Ap\&SS, 351, 1

\bibitem[{Pastorello {et~al.}(2013)Pastorello, Cappellaro, Inserra, Smartt,
  Pignata, Benetti, Valenti, Fraser, Tak{\'a}ts, Benitez, Botticella,
  Brimacombe, Bufano, Cellier-Holzem, Costado, Cupani, Curtis, Elias-Rosa,
  Ergon, Fynbo, Hambsch, Hamuy, Harutyunyan, Ivarson, Kankare, Martin, Kotak,
  LaCluyze, Maguire, Mattila, Maza, McCrum, Miluzio, Norgaard-Nielsen,
  Nysewander, Ochner, Pan, Pumo, Reichart, Tan, Taubenberger, Tomasella,
  Turatto, \& Wright}]{pastorello13}
Pastorello, A., Cappellaro, E., Inserra, C., {et~al.} 2013, ApJ, 767, 1

\bibitem[{Phillips(1993)}]{phillips93}
Phillips, M.~M. 1993, ApJ, 413, L105

\bibitem[{Prieto {et~al.}(2013)Prieto, Brimacombe, Drake, \&
  Howerton}]{prieto13}
Prieto, J.~L., Brimacombe, J., Drake, A.~J., \& Howerton, S. 2013, ApJL, 763,
  L27

\bibitem[{Prieto {et~al.}(2007)Prieto, Garnavich, Phillips, DePoy, Parrent,
  Pooley, Dwarkadas, Baron, Bassett, Becker, Cinabro, DeJongh, Dilday, Doi,
  Frieman, Hogan, Holtzman, Jha, Kessler, Konishi, Lampeitl, Marriner,
  Marshall, Miknaitis, Nichol, Riess, Richmond, Romani, Sako, Schneider, Smith,
  Takanashi, Tokita, van~der Heyden, Yasuda, Zheng, Wheeler, Barentine,
  Dembicky, Eastman, Frank, Ketzeback, McMillan, Morrell, Folatelli, Contreras,
  Burns, Freedman, Gonzalez, Hamuy, Krzeminski, Madore, Murphy, Persson, Roth,
  \& Suntzeff}]{prieto07}
Prieto, J.~L., Garnavich, P.~M., Phillips, M.~M., {et~al.} 2007, eprint arXiv,
  0706, 4088, 63 pages, 16 figures, submitted to AJ

\bibitem[{Raskin {et~al.}(2014)Raskin, Kasen, Moll, Schwab, \&
  Woosley}]{raskin14}
Raskin, C., Kasen, D., Moll, R., Schwab, J., \& Woosley, S. 2014, ApJ, 788, 75

\bibitem[{Schlegel(1990)}]{schlegel90}
Schlegel, E.~M. 1990, MNRAS, 244, 269

\bibitem[{Schmidt {et~al.}(1989)Schmidt, Weymann, \& Foltz}]{schmidt89}
Schmidt, G.~D., Weymann, R.~J., \& Foltz, C.~B. 1989, PASP, 101, 713

\bibitem[{Silverman {et~al.}(2011)Silverman, Ganeshalingam, Li, Filippenko,
  Miller, \& Poznanski}]{silverman11}
Silverman, J.~M., Ganeshalingam, M., Li, W., {et~al.} 2011, MNRAS, 410, 585

\bibitem[{Silverman {et~al.}(2012{\natexlab{a}})Silverman, Kong, \&
  Filippenko}]{silverman12bsnip2}
Silverman, J.~M., Kong, J.~J., \& Filippenko, A.~V. 2012{\natexlab{a}}, MNRAS,
  425, 1819

\bibitem[{Silverman {et~al.}(2012{\natexlab{b}})Silverman, Foley, Filippenko,
  Ganeshalingam, Barth, Chornock, Griffith, Kong, Lee, Leonard, Matheson,
  Miller, Steele, Barris, Bloom, Cobb, Coil, Desroches, Gates, Ho, Jha,
  Kandrashoff, Li, Mandel, Modjaz, Moore, Mostardi, Papenkova, Park, Perley,
  Poznanski, Reuter, Scala, Serduke, Shields, Swift, Tonry, van Dyk, Wang, \&
  Wong}]{silverman12bsnip1}
Silverman, J.~M., Foley, R.~J., Filippenko, A.~V., {et~al.} 2012{\natexlab{b}},
  MNRAS, 425, 1789

\bibitem[{Silverman {et~al.}(2013{\natexlab{a}})Silverman, Nugent, Gal-Yam,
  Sullivan, Howell, Filippenko, Pan, Cenko, \& Hook}]{silverman13a}
Silverman, J.~M., Nugent, P.~E., Gal-Yam, A., {et~al.} 2013{\natexlab{a}}, ApJ,
  772, 125

\bibitem[{Silverman {et~al.}(2013{\natexlab{b}})Silverman, Nugent, Gal-Yam,
  Sullivan, Howell, Filippenko, Arcavi, Ben-Ami, Bloom, Cenko, Cao, Chornock,
  Clubb, Coil, Foley, Graham, Griffith, Horesh, Kasliwal, Kulkarni, Leonard,
  Li, Matheson, Miller, Modjaz, Ofek, Pan, Perley, Poznanski, Quimby, Steele,
  Sternberg, Xu, \& Yaron}]{silverman13b}
---. 2013{\natexlab{b}}, ApJS, 207, 3

\bibitem[{Simcoe {et~al.}(2008)Simcoe, Burgasser, Bernstein, Bigelow, Fishner,
  Forrest, McMurtry, Pipher, Schechter, \& Smith}]{simcoe08}
Simcoe, R.~A., Burgasser, A.~J., Bernstein, R.~A., {et~al.} 2008, Proc. of
  SPIE, 7014, 0

\bibitem[{Skrutskie {et~al.}(2006)Skrutskie, Cutri, Stiening, Weinberg,
  Schneider, Carpenter, Beichman, Capps, Chester, Elias, Huchra, Liebert,
  Lonsdale, Monet, Price, Seitzer, Jarrett, Kirkpatrick, Gizis, Howard, Evans,
  Fowler, Fullmer, Hurt, Light, Kopan, Marsh, McCallon, Tam, Dyk, \&
  Wheelock}]{skrutskie06}
Skrutskie, M.~F., Cutri, R.~M., Stiening, R., {et~al.} 2006, AJ, 131, 1163

\bibitem[{Smith {et~al.}(2013)Smith, Mauerhan, Kasliwal, \&
  Burgasser}]{smith13ip}
Smith, N., Mauerhan, J.~C., Kasliwal, M.~M., \& Burgasser, A.~J. 2013, MNRAS,
  434, 2721

\bibitem[{Smith {et~al.}(2014)Smith, Mauerhan, \& Prieto}]{smith14ip}
Smith, N., Mauerhan, J.~C., \& Prieto, J.~L. 2014, MNRAS, 438, 1191

\bibitem[{Smith {et~al.}(2012)Smith, Mauerhan, Silverman, Ganeshalingam,
  Filippenko, Cenko, Clubb, \& Kandrashoff}]{smith12hw}
Smith, N., Mauerhan, J.~C., Silverman, J.~M., {et~al.} 2012, MNRAS, 426, 1905

\bibitem[{Smith \& Tombleson(2014)}]{smith14lbv}
Smith, N., \& Tombleson, R. 2014, arXiv: 1406.7431

\bibitem[{Smith {et~al.}(2009)Smith, Silverman, Chornock, Filippenko, Wang, Li,
  Ganeshalingam, Foley, Rex, \& Steele}]{smith09ip}
Smith, N., Silverman, J.~M., Chornock, R., {et~al.} 2009, ApJ, 695, 1334

\bibitem[{Smith {et~al.}(2010)Smith, Miller, Li, Filippenko, Silverman, Howard,
  Nugent, Marcy, Bloom, Ghez, Lu, Yelda, Bernstein, \& Colucci}]{smith10ip}
Smith, N., Miller, A., Li, W., {et~al.} 2010, AJ, 139, 1451

\bibitem[{Taddia {et~al.}(2012)Taddia, Stritzinger, Phillips, Burns,
  Heinrich-Josties, Morrell, Sollerman, Valenti, Anderson, Boldt, Campillay,
  Castellon, Contreras, Folatelli, Freedman, Hamuy, Krzeminski, Leloudas,
  Maeda, Persson, Roth, \& Suntzeff}]{taddia12}
Taddia, F., Stritzinger, M.~D., Phillips, M.~M., {et~al.} 2012, A{\&}A, 545, L7

\bibitem[{Taubenberger {et~al.}(2011)Taubenberger, Benetti, Childress, Pakmor,
  Hachinger, Mazzali, Stanishev, Elias-Rosa, Agnoletto, Bufano, Ergon,
  Harutyunyan, Inserra, Kankare, Kromer, Navasardyan, Nicolas, Pastorello,
  Prosperi, Salgado, Sollerman, Stritzinger, Turatto, Valenti, \&
  Hillebrandt}]{taubenberger11}
Taubenberger, S., Benetti, S., Childress, M., {et~al.} 2011, MNRAS, 412, 2735

\bibitem[{Taubenberger {et~al.}(2013)Taubenberger, Kromer, Hachinger, Mazzali,
  Benetti, Nugent, Scalzo, Pakmor, Stanishev, Spyromilio, Bufano, Sim,
  Leibundgut, \& Hillebrandt}]{taubenberger13}
Taubenberger, S., Kromer, M., Hachinger, S., {et~al.} 2013, MNRAS, 432, 3117

\bibitem[{Turatto {et~al.}(1993)Turatto, Cappellaro, Danziger, Benetti,
  Gouiffes, \& Valle}]{turatto93}
Turatto, M., Cappellaro, E., Danziger, I.~J., {et~al.} 1993, MNRAS, 262, 128

\bibitem[{Turatto {et~al.}(2000)Turatto, Mazzali, Suzuki, Young, Nomoto,
  Benetti, Cappellaro, Danziger, \& Patat}]{turatto00}
Turatto, M., Mazzali, P.~A., Suzuki, T., {et~al.} 2000, Supernovae and
  gamma-ray bursts: The Greatest Explosions Since the Big Bang: poster papers
  from the Space Telescope Science Institute Symposium, 72

\bibitem[{Vacca {et~al.}(2003)Vacca, Cushing, \& Rayner}]{vacca03}
Vacca, W.~D., Cushing, M.~C., \& Rayner, J.~T. 2003, The Publications of the
  Astronomical Society of the Pacific, 115, 389

\bibitem[{Valenti {et~al.}(2012)Valenti, Pastorello, Cappellaro,
  Benitez-Herrera, Taubenberger, Smartt, Young, Sullivan, Gal-Yam, Yaron,
  Baltay, Ellman, Hadjiyska, McKinnon, Rabinowitz, Feindt, Kowalski, Nugent, \&
  Pignata}]{valenti12}
Valenti, S., Pastorello, A., Cappellaro, E., {et~al.} 2012, ATEL, 4076, 1

\bibitem[{van Marle {et~al.}(2010)van Marle, Smith, Owocki, \& van
  Veelen}]{marle10}
van Marle, A.~J., Smith, N., Owocki, S.~P., \& van Veelen, B. 2010, MNRAS, 407,
  2305

\bibitem[{Wade \& Horne(1988)}]{wade88}
Wade, R.~A., \& Horne, K. 1988, ApJ, 324, 411

\bibitem[{Wagner \& House(1968)}]{wagner68}
Wagner, W.~J., \& House, L.~L. 1968, Solar Physics, 5, 55

\bibitem[{Watson {et~al.}(2012)Watson, Richer, Bloom, Butler, Cese{\~n}a,
  Clark, Colorado, C{\'o}rdova, Farah, Fox-Machado, Fox, Garc{\'\i}a, Georgiev,
  Gonz{\'a}lez, Guisa, Guti{\'e}rrez, Herrera, Klein, Kutyrev, Lazo, Lee,
  L{\'o}pez, Luna, Mart{\'\i}nez, Murillo, Murillo, N{\'u}{\~n}ez, Prochaska,
  Ochoa, Quir{\'o}s, Rapchun, Rom{\'a}n-Z{\'u}{\~n}iga, \& Valyavin}]{watson12}
Watson, A.~M., Richer, M.~G., Bloom, J.~S., {et~al.} 2012, Proc. of SPIE, 8444,
  doi:10.1117/12.926927

\bibitem[{Williams {et~al.}(1991)Williams, Hamuy, Phillips, Heathcote, Wells,
  \& Navarrete}]{williams91}
Williams, R.~E., Hamuy, M., Phillips, M.~M., {et~al.} 1991, ApJ, 376, 721

\bibitem[{Wilson {et~al.}(2004)Wilson, Henderson, Herter, Matthews, Skrutskie,
  Adams, Moon, Smith, Gautier, Ressler, Soifer, Lin, Howard, LaMarr, Stolberg,
  \& Zink}]{wilson04}
Wilson, J.~C., Henderson, C.~P., Herter, T.~L., {et~al.} 2004, Proc. of SPIE,
  5492, 1295

\bibitem[{Wood-Vasey {et~al.}(2004)Wood-Vasey, Wang, \&
  Aldering}]{wood-vasey04}
Wood-Vasey, W.~M., Wang, L., \& Aldering, G. 2004, ApJ, 616, 339

\bibitem[{Yaron \& Gal-Yam(2012)}]{yaron12}
Yaron, O., \& Gal-Yam, A. 2012, PASP, 124, 668

\end{thebibliography}

\end{document}